\documentclass[aip,
 amsmath,amssymb,
 reprint,%
]{revtex4-1}

\usepackage{graphicx}% Include figure files
\usepackage{dcolumn}% Align table columns on decimal point
\usepackage{bm}% bold math
\usepackage[utf8]{inputenc}
\usepackage[T1]{fontenc}
\usepackage{mathptmx}
\usepackage{etoolbox}
\usepackage{color}

\usepackage[hidelinks]{hyperref}
\hypersetup{
    colorlinks = true,
    citecolor = blue,
    linkcolor = blue,
    urlcolor = blue
}

\makeatletter
\def\@email#1#2{%
 \endgroup
 \patchcmd{\titleblock@produce}
  {\frontmatter@RRAPformat}
  {\frontmatter@RRAPformat{\produce@RRAP{*#1\href{mailto:#2}{#2}}}\frontmatter@RRAPformat}
  {}{}
}%
\makeatother
\begin{document}

\title{Heterogeneous response and non-Markovianity in the microrheology of semisolid viscoelastic materials}
\author{T. N. Azevedo}
\author{L. G. Rizzi$^{*,}$}%
\email{lerizzi@ufv.br.}
\affiliation{ 
Departamento de F\'isica, Universidade Federal de Vi\c{c}osa (UFV), Av. P. H. Rolfs, s/n, 36570-900, Vi\c{c}osa, Brazil.
}%

\date{\today}

\begin{abstract}
	Recent works indicate that heterogeneous response and non-Markovianity may yield recognizable hallmarks in the microrheology of semisolid viscoelastic materials.
	Here we perform numerical simulations using a non-Markovian overdamped Langevin approach to explore how the microrheology experienced by probe particles immersed in an effective semisolid material can be influenced by its micro-heterogeneities.
	Our results show that, besides affecting the mean squared displacement, the time-dependent diffusion coefficient, and the shear moduli, the micro-heterogeneities lead to displacement distributions that deviate from the usual Gaussian behavior.
	In addition, our study provides an analytical way to characterize the micro-heterogeneities of semisolid viscoelastic materials through their microrheology.
\end{abstract}

\maketitle

\section{Introduction}

	Although traditional approaches based on the Kelvin Voigt (KV) and Maxwell models~\cite{ferrybook,raobook} have been widely considered in the literature to explain viscoelasticity in isotropic materials qualitatively, experimental results usually display microrheological features that cannot be described by their simple expressions.
	The KV model, in particular, which is commonly used to illustrate the viscoelastic behavior of semisolids, is mechanically characterized by both a constant storage modulus $G^{\prime}(\omega) = G_0$ and a constant viscosity $\eta_0=G^{\prime \prime}(\omega)/\omega$, which implies that its loss modulus $G^{\prime \prime}(\omega)$ is linear.
	From the microrheology point-of-view, one has that the overdamped Brownian dynamics of spherical probe particles with radius $a$ immersed in an isotropic viscoelastic material can be generally related to its compliance $J(\tau)$ as~\cite{Rizzi_Tassieri_2018}
\begin{equation}
J(\tau) = \frac{ 3 \pi a}{d_e k_BT} \langle \Delta r^2(\tau) \rangle~~,
\label{GSER-Jtau-MSD}
\end{equation}
where $\langle \Delta r^2(\tau) \rangle$ is the mean squared displacement (MSD) computed from the random walk of the probe particles in $d_e$ Euclidean dimensions, $k_B$ is the Boltzmann's constant, and $T$ is the absolute temperature of the material.
	Additionally, by considering the relation~\cite{Rizzi_Tassieri_2018} $G^{*}(\omega)=1/\mathbf{i}\omega \hat{J}(\omega)$, with $\mathbf{i}=\sqrt{-1}$, one can evaluate the storage and loss moduli that define the complex modulus $G^{*}(\omega) = G^{\prime}(\omega) + \mathbf{i}G^{\prime \prime}(\omega)$ in terms of the Fourier transform of the compliance $\hat{J}(\omega)=\mathcal{F}[J(\tau);\omega]$.
	Hence, one can determine the MSD of probe particles immersed in a semisolid KV material as~\cite{azevedo2025softmatter,azevedo2020jphysconfser}
\begin{equation}
\langle \Delta r^2(\tau) \rangle = 2 \frac{d_e k_BT}{\kappa} \left( 1 - e^{- (\kappa/\zeta) \tau} \right)~~,
\label{MSD-KV-usual}
\end{equation}
where $\zeta=6 \pi a \eta_0$ and $\kappa$ denote the elastic constant of the effective spring that traps the probe particle in a limited region of the sample. 
	Indeed, from Eq.~\ref{MSD-KV-usual}, one has that $\lim_{\tau \rightarrow \infty} \langle \Delta r^2(\tau) \rangle = 2 d_e k_BT/\kappa$, so Eq.~\ref{GSER-Jtau-MSD} can be used to demonstrate that $\kappa$ is directly related to the plateau modulus of the semisolid as $G_0 = \lim_{\omega \rightarrow 0} G^{\prime}(\omega) = \kappa/(6 \pi a)$.

\begin{figure*}
\centering
\includegraphics[width=0.85\textwidth]{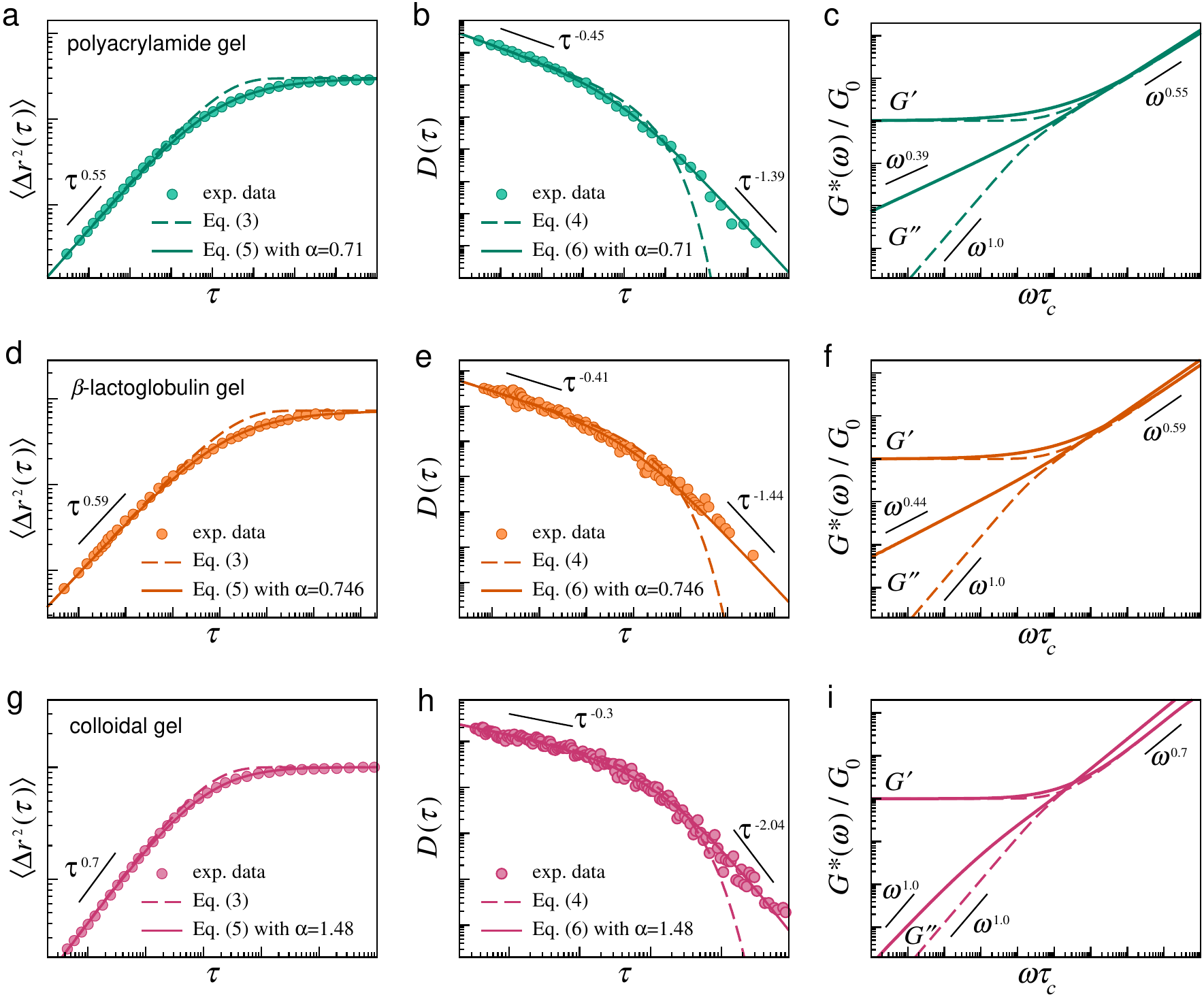}
\caption{Examples of experimental data (filled circles) illustrating the hallmarks of heterogeneous response and non-Markovianity in semisolid viscoelastic materials: mean squared displacement $\langle \Delta r^2(\tau) \rangle$, time-dependent diffusion coefficient $D(\tau)$, and the shear moduli $G^{\prime}(\omega)$ and $G^{\prime \prime}(\omega)$.
	Panels (a)-(c) include data extracted from Ref.~\cite{furst2008prl} and correspond to the master curves obtained for polyacrylamide gels, while (d)-(f) include data obtained for $\beta$-lactoglobulin-based gels extracted from Ref.~\cite{donald2009eurphysJE}, and (g)-(i) correspond to data on colloidal gels extracted from Ref.~\cite{romer2014epl}.
	Dashed and continuous lines denote results obtained from fits of the experimental data to Eqs.~\ref{MSD-NM-KV} and~\ref{difcoefNM-KV} (i.e., NM-KV model), and Eqs.~\ref{MSDjrheol2020} and~\ref{Dtaurheol2020} (NM-KVMH model, see Ref.~\cite{rizzi2020jrheol}), respectively.
	The shear moduli $G^{\prime}(\omega)$ and $G^{\prime \prime}(\omega)$ were obtained numerically from the MSD curves through the GSER given by Eq.~\ref{GSER-Jtau-MSD} using the numerical procedure described in Ref.~\cite{Evans2009}.
	Physical units are omitted since the master curves were used only to illustrate the behavior of these quantities (we referred the reader to the original articles for further information).
}
\label{examples}
\end{figure*}

	As shown in the examples included in Fig.~\ref{examples}, the microrheology of semisolids is usually different from the ideal KV behavior.
	Although the experimentally obtained MSDs display a plateau at later times as the KV model, they present a power-law behavior at short times where $\langle \Delta r^2(\tau) \rangle \propto \tau^{n}$ with $n<1$, which is markedly different from the microrheology predicted by Eq.~\ref{MSD-KV-usual}, i.e., $\langle \Delta r^2(\tau) \rangle \propto \tau$ for $\tau \ll \zeta/\kappa$.
	As discussed in Ref.~\cite{rizzi2020jrheol}, exponents $n$ that are less than one can be attributed to the memory effects related to the non-Markovian dynamics of the effective structures that are present in the viscoelastic material, such as polymer chains and protein fibers~\cite{jahnel2008softmatter,georgiades2013biopolymers}.
	For instance, it is demonstrated in Ref.~\cite{duarte2021sm} that flexible structures like Rouse chains would lead to exponents equal to $n=0.5$, while semi-flexible structures like collagen fibers and DNA filaments would lead the exponent to be closer to $n=0.75$.
	With a few exceptions of viscoelastic materials that display nearly ideal KV behavior~\cite{azevedo2025softmatter}, most semisolids present a microrheology as the examples in Fig.~\ref{examples} and exhibit exponents $n$ in that range (see, e.g., Refs.~\cite{romer2000prl,veerman2006macromol,furst2008prl,larsen2008koreaust,larsen2009macromol,donald2009eurphysJE,donald2009langmuir,schultz2012softmatter,donald2012softmatter,donald2014eurphysJE,romer2014epl}).
	Nevertheless, it is worth noting that less-than-one exponents characterizing sub-diffusive dynamics of probe particles can be found for porous materials, but that is attributed to the fact that the radii of the particles are smaller than the structures in these materials~\cite{BenAvraham2002}.
	In such situations, it is usually acknowledged that the MSD may not be taken as a probe of the microrheology of the material, since Eq.~\ref{GSER-Jtau-MSD} would give a compliance that cannot be comparable to the one obtained from bulk rheology~\cite{Rizzi_Tassieri_2018,Waigh_2016}.

%%%%%%%%%%%%%%%%%%%%%%%%%%%%%%%%%%%%%%%%%
%%%%%%%%%%%% FIG. 1: EXAMPLES %%%%%%%%%%%
%%%%%%%%%%%%%%%%%%%%%%%%%%%%%%%%%%%%%%%%%
%%%%%%%%%%%%%%%%%%%%%%%%%%%%%%%%%%%%%%%%%
%%%%%%%%%%%%%%%%%%%%%%%%%%%%%%%%%%%%%%%%%
%%%%%%%%%%%%%%%%%%%%%%%%%%%%%%%%%%%%%%%%%

	Alternatively, several analyses of experimentally determined MSDs have been carried out through a generalization of the KV model that is defined by a slightly different expression, which is given by
\begin{equation}
\langle \Delta r^2(\tau) \rangle = 2 \frac{d_e k_BT}{\kappa} \left\{ 1 - \exp\left[ -\left(\frac{\tau}{\tau_c} \right)^{n}  \, 
\right] \right\}~~,
\label{MSD-NM-KV}
\end{equation}
where $\tau_c$ corresponds to a mean characteristic time that is analog to the characteristic time $\zeta/\kappa$ of the usual KV model.
	Arguably, since Eq.~\ref{MSD-NM-KV} yields $\langle \Delta r^2(\tau) \rangle \propto \tau^n$ at short times, and this power-law behavior is due to the non-Markovian character of the stochastic dynamics of the structures in the material, we refer to this expression as the MSD of the non-Markovian Kelvin Voigt (NM-KV) model.
	Equation~\ref{MSD-NM-KV} has been used to fit the MSD curves obtained from several semisolid materials, including gels based on proteins~\cite{rizzi2017jcp}, colloidal particles~\cite{KRALL199719,krall1998prl,romer2000prl,romer2014epl}, and crosslinked polymers~\cite{teixeira2007jphyschemB}.
	As shown in Fig.~\ref{examples}, it yields acceptable fits both at short and later times, with
discrepancies observed mainly at intermediate times.
	In fact, a careful analysis indicates that the NM-KV model has limited applicability, and that can be realized from the time-dependent diffusion coefficient $D(\tau)$, which is evaluated from the derivative of Eq.~\ref{MSD-NM-KV}, that is,
\begin{equation}
D(\tau) 
 = \frac{1}{2 d_e} \frac{d \langle \Delta r^2(\tau) \rangle}{d\tau} 
 = \frac{n \, k_BT}{ \kappa \tau_c} \, \left( \frac{\tau}{\tau_c} \right)^{n-1} \, \exp \left[ -\left( \frac{\tau}{\tau_c} \right)^{n} \, \right] ~~.
\label{difcoefNM-KV}   
\end{equation}
	From Fig.~\ref{examples} one observes that the experimental curves match the short-time power-law behavior expected from this expression, i.e., $D(\tau) \propto \tau^{n-1}$, but the data clearly deviate from the exponential behavior predicted by Eq.~\ref{difcoefNM-KV} at later times.

	More recently, by considering a non-Markovian approach based on an overdamped generalized Langevin equation (GLE), Ref.~\cite{rizzi2020jrheol} established that the MSD of probe particles immersed in isotropic viscoelastic materials can be better described by an expression that is given by
\begin{equation}
\langle \Delta r^2(\tau) \rangle
 = 2 \frac{d_e k_BT}{\kappa} \left\{ 1 - \left[ \frac{1}{\alpha}  \left(\frac{\tau}{\tau_c} \right)^{n} + 1 \right]^{-\alpha} \right\} ~~,
\label{MSDjrheol2020} 
\end{equation}
where the exponent $\alpha$ is a parameter that characterizes the distribution of microrheological properties (e.g., elastic constants and drag coefficients) at different mesoscopic regions of the sample (see also Ref.~\cite{azevedo2025softmatter}).
	A few remarks are due here, but the first thing to note is that the above expression gives a more suitable description of the experimental data presented in Fig.~\ref{examples}. 
	Not only the data obtained for the MSDs are better fitted to Eq.~\ref{MSDjrheol2020} but also the results obtained for the time-dependent diffusion coefficient, which expression is evaluated from the above MSD and reads as~\cite{rizzi2020jrheol}
\begin{equation}
D(\tau) = \frac{n \, k_BT}{\kappa \tau_c} \, \left( \frac{\tau}{\tau_c} \right)^{n-1} \Bigg/  \left[  \frac{1}{\alpha}  \left( \frac{\tau}{\tau_c} \right)^{n} + 1 \right]^{1+\alpha} ~~.
\label{Dtaurheol2020}
\end{equation}
	Accordingly, the comparisons included in Fig.~\ref{examples} indicate that the data extracted for the time-dependent diffusion coefficient $D(\tau)$ for these different semisolids can be described by Eq.~\ref{Dtaurheol2020} at both short times, where $D(\tau) \propto \tau^{n-1}$, and at later times, where $D(\tau) \propto \tau^{-(1+\alpha n)}$.
    The later time behavior of the time-dependent diffusion coefficient $D(\tau)$ described by Eq.~\ref{Dtaurheol2020}, which is clearly not exponential, corresponds to the hallmark of a heterogeneous viscoelastic response~\cite{rizzi2020jrheol}.
	In addition to the examples included in Fig.~\ref{examples}, Ref.~\cite{teixeira2007jphyschemB} shows that there are many gels that display such a power-law behavior at later times.

	As detailed in Ref.~\cite{rizzi2020jrheol}, the exponent $\alpha$ characterizes a generalized gamma distribution~\cite{Crooks2019}, which was used to incorporate the effect of micro-heterogeneities into the MSD, in a way that different probe particles at different mesoscopic regions of the sample are subjected to slightly different viscoelastic properties.
	The use of a gamma distribution was originally proposed because of the relationship between elastic constants and polymer sizes~\cite{zaccone2014jrheol}, as the gamma distributions for the latter would also imply the same distribution for the former.
	Despite that, Ref.~\cite{azevedo2025softmatter} included a direct comparison between experimentally obtained displacement (i.e., van Hove) distributions and numerical results, further validating the use of gamma distributions to model the heterogeneous response of semisolid viscoelastic materials.
	Also, the one-parameter generalized gamma distribution is a convenient choice since the limit of large $\alpha$ corresponds to the limit of a homogeneous material, where the terms inside the brackets in Eqs.~\ref{MSDjrheol2020} and~\ref{Dtaurheol2020} retrieve stretched exponentials.
	Hence, these two expressions can be interpreted as generalizations of the expressions of the NM-KV, i.e., Eqs.~\ref{MSD-NM-KV} and~\ref{difcoefNM-KV}.
	In that way, one can refer to the model introduced in Ref.~\cite{rizzi2020jrheol}, i.e., the model that yields Eqs.~\ref{MSDjrheol2020} and~\ref{Dtaurheol2020}, as a non-Markovian Kelvin Voigt model with micro-heterogeneities (NM-KVMH).

	In terms of viscoelastic response, Fig.~\ref{examples} indicates that the hallmark of non-Markovianity corresponds to the power-law behavior observed in the shear moduli at high frequencies, where~\cite{rizzi2020jrheol} $G^{\prime}(\omega) \propto G^{\prime \prime}(\omega) \propto \omega^{n}$.
	On the other hand, the presence of micro-heterogeneities seems to smooth the transition between the plateau and the power-law regimes observed for the storage modulus $G^{\prime}(\omega)$ at intermediate frequencies.
	The microrheology obtained from MSDs like those described by Eq.~\ref{MSDjrheol2020} indicates that the loss modulus $G^{\prime \prime}(\omega)$ should be also affected by the presence of micro-heterogeneities, mainly at low frequencies, where the NM-KVMH model predicts~\cite{rizzi2020jrheol} $G^{\prime \prime}(\omega) \propto \omega^{\alpha n}$, while the NM-KV expression, Eq.~\ref{MSD-NM-KV}, yields a linear behavior like the usual KV model, i.e., $G^{\prime \prime}(\omega) \propto \omega$ (see Fig.~\ref{examples}).

	In this work, we present a strategy and include numerical results obtained from Brownian simulations to further validate the fundamental ideas of Refs.~\cite{rizzi2020jrheol,azevedo2025softmatter} underpinning the non-Markovian and heterogeneous response of isotropic semisolid viscoelastic materials.
	The manuscript is organized as follows.
	In Sec.~\ref{approxNM-KVMH} we revise the conceptual framework on how one can take into account micro-heterogeneities to describe the viscoelastic response of semisolids.
	In particular, we introduce an approximation to the NM-KVMH in terms of the Prony series which allows one to perform non-Markovian Brownian simulations, which are, in turn, described in Sec.~\ref{GLEsimulations}.
	Results obtained for the MSD, time-dependent diffusion coefficient, displacement distributions, and shear moduli are presented in Sec.~\ref{results}, while Sec.~\ref{conclusions} includes our concluding remarks.

\section{Micro-heterogeneities and the approximated NM-KVMH model}
\label{approxNM-KVMH}

	Based on the ideas of Ref.~\cite{rizzi2020jrheol} and following the approach used to obtain the Markovian KVMH model~\cite{azevedo2025softmatter}, we assume that the MSD can be evaluated as the average over trajectories where different probe particles experience different viscoelastic properties due to micro-heterogeneities, that is,
\begin{equation}
\langle \Delta x^2 (\tau) \rangle = \int_{0}^{\infty} \langle \Delta x^2 (\tau) \rangle_{\xi} \, \rho(\xi) \, d\xi~~.
\label{MSD-rhoxi}
\end{equation}
	As detailed in Ref.~\cite{azevedo2025softmatter}, the locally defined microrheological properties of mesoscopic regions of the system can be incorporated into the model through a region-dependent variable $\xi=\xi(\vec{R})$, which is distributed according to a generalized gamma distribution~\cite{Crooks2019}
\begin{equation}
\rho(\xi) = \frac{ \xi^{-(1-p)} \, e^{-\xi} }{ \Gamma(p) }~~,
\label{distribution-rho-xi}
\end{equation}
with $\Gamma(p)$ being the usual gamma function.

	The idea here is that, by considering the local MSD $\langle \Delta x^2 (\tau) \rangle_{\xi}$ that is given by the NM-KV model, Eq.~\ref{MSD-NM-KV}, one can use Eq.~\ref{MSD-rhoxi} to take into account the micro-heterogeneities and obtain the MSD of the NM-KVMH model given by Eq.~\ref{MSDjrheol2020} (as in Ref.~\cite{rizzi2020jrheol}).
	As it happens, in order to incorporate non-Markovian effects through memory kernels, one needs a reliable (and not so cumbersome) numerical method.
	Although it might not be impossible to set a memory kernel to obtain a local MSD that is given by the NM-KV model, Eq.~\ref{MSD-NM-KV}, it is not easy to find suitable methods that can be used to eventually simulate non-Markovian processes for such an arbitrary memory kernel.
	Hence, to approximate the NM-KV model and perform the numerical simulations, we assume instead that the local MSD given by Eq.~\ref{MSD-NM-KV} can be effectively approximated by a finite Prony series and is expressed as
\begin{equation}
\langle \Delta x^2(\tau) \rangle_{\xi} = 2 \frac{k_BT }{\kappa} - \sum_{j=1}^{N+1} q_{j,\xi} \exp \left(-\tau/\gamma_{j,\xi} \right)~~.
\label{MSD-NM-KVMH-local}
\end{equation}
	Here, just like in the subsequent sections, we assume $d_e=1$ Euclidean dimensions without loss of generality and for the sake of computational efficiency.
	The parameters $q_{j,\xi}$ and $\gamma_{j,\xi}$ in Eq.~\ref{MSD-NM-KVMH-local} are used to define the local microrheological properties of the viscoelastic semisolid material.
	While the continuous index $\xi$ denotes a specific region of the sample and is used to describe the heterogeneities within the material, the index $j$ is used to indicate a summation over $N+1$ relaxation modes and should incorporate the non-Markovianity of the response.
	To illustrate that, we include in Table~\ref{tabela-referencia} reference parameters $q_j^{*}$ and $\gamma_{j}^{*}$ which lead the MSD given by the Prony series, Eq.~\ref{MSD-NM-KVMH-local}, to be a reasonably good approximation to the MSD given by the NM-KV, Eq.~\ref{MSD-NM-KV}, with an exponent equal to $n=0.5$ (as discussed in Ref.~\cite{mauro2018physicaA}, other exponents could be considered).
	The inspection of the approximation indicates that it works only for a restricted temporal window, i.e., $\tau \geq 0.1\,$s, but that includes the regions of interest where the MSD displays the power-law and the plateau regimes.

\begin{table}
\caption{
	Reference parameters $q_j^{*}$ and $\gamma_j^{*}$ used to approximate the MSD of the NM-KV model, Eq.~\ref{MSD-NM-KV} with exponent $n=0.5$, by a Prony series, Eq.~\ref{MSD-NM-KVMH-local}, with $N=7$ (i.e., eight modes). These parameters, in particular, yield the reference parameters $\mu_0^{*}$, $\{ c_j^{*} \}$, and  $\{ \Lambda_j^{*} \}$, and serve as input to obtain the parameters $\mu_{0,m}$, $\{ c_{j,m} \}$, and  $\{ \Lambda_{j,m} \}$ used in the GLE-based simulations, Eqs.~\ref{eq:x-update-scheme} and~\ref{eq:q-update-scheme}.
}
\label{tabela-referencia}
\begin{ruledtabular}
\begin{tabular}{ccccc}
$j$ & $q_j^{*}$ [$10^{-4}\,\mu$m$^2$] & $\gamma_j^{*}$ [$s$] & $c_j^{*}$ [kg$^{-1}$] & $\Lambda_j^{*}$ [$s$]\\
\hline
1 & 1.3629992176 & 12.92824 &   0.0025456721    & 10.7685962 \\
2 & 5.5924072515 & 5.25348 &    0.0196891753    &  3.36553605 \\
3 & 7.9266388980 & 2.15936 &    0.1084346927    &  1.1299373 \\
4 & 7.3722725229 & 0.85266 &    0.6194283423    &  0.3722605 \\
5 & 5.6189615464 & 0.31087 &    4.1543978326    &  0.1125437 \\
6 & 3.8296385810 & 0.10048 &   37.8758355209    &  0.0288819 \\
7 & 2.4284448297 & 0.02694 & 1366.8002874243    &  0.0045039 \\
8 & 2.2436560380 & 0.00423 &                    &  
\end{tabular}
\end{ruledtabular}
\end{table}

	Given the reference parameters of Table~\ref{tabela-referencia}, one can obtain rescaled parameters that describe the micro-heterogeneities of the mesoscopic regions of the material as in Ref.~\cite{azevedo2025softmatter}.
	In particular, we consider that $q_{j,\xi} = q_{j}^{*}$ and that the local drag coefficients are given so that the characteristic times scale as
\begin{equation}
\gamma_{j,\xi} = \frac{p}{\xi} \gamma_j^{*}~~,
\label{local-gamma-xi}
\end{equation}
where $\xi$ is distributed according to Eq.~\ref{distribution-rho-xi}.
	Hence, by inserting Eq.~\ref{MSD-NM-KVMH-local} into Eq.~\ref{MSD-rhoxi} and evaluating the corresponding integrals~\cite{gradshteyn} with the distribution $\rho(\xi)$ given by Eq.~\ref{distribution-rho-xi}, one obtains the following expression for the MSD
\begin{equation}
\langle \Delta x^2(\tau) \rangle = 2 \frac{k_BT}{\kappa} - \sum_{j=1}^{N+1}q_j^{*}\left[ 1 + \frac{1}{p} \frac{\tau}{\gamma_j^{*}} \right]^{-p}~~.
\label{MSD-NM-KVMH-final}
\end{equation}
	Additionally, the time-dependent diffusion coefficient $D(\tau)$ can be evaluated through the derivative of this MSD and is given by
\begin{equation}
D(\tau) = \frac{1}{2} \sum_{j=1}^{N+1} \frac{q_j^{*}}{\gamma_j^{*}} 
\left[ 1 + \frac{1}{p} \frac{\tau}{\gamma_j^{*}}  \right]^{-(1+p)} ~~.
\label{Dtau-NM-KVMH-final}
\end{equation}
	One should note that, just as in the NM-KVMH, the terms inside brackets in the above expressions retrieve exponentials for large values of $p$, and it would recover the exponentials as in Eq.~\ref{MSD-NM-KVMH-local} so that the semisolid material is considered homogeneous.
	In addition, it is worth emphasizing that, although Eq.~\ref{MSD-NM-KVMH-local} and its derivative are, respectively, equivalent to Eqs.~\ref{MSD-NM-KV} and~\ref{difcoefNM-KV} of the NM-KV model (within the range $\tau \geq 0.1\,$s for the parameters listed in Table~\ref{tabela-referencia}), the above expressions are not strictly equivalent to the expressions of the NM-KVMH, i.e., Eqs.~\ref{MSDjrheol2020} and~\ref{Dtaurheol2020}.
	Besides, it should be clear that the limit $\tau \rightarrow 0$ given by the above expression is different from the expressions obtained from those two non-Markovian models since it yields a finite value $\lim_{\tau \rightarrow 0}D(\tau) = (1/2) \sum_{j=1}^{N+1} \left( q_j^{*}/ \gamma_j^{*}\right) > 0$ instead of a power-law behavior, i.e., $D(\tau) \propto \tau^{n-1}$.

	Next, we use the fact that the MSD of probe particles trapped in a particular region of the sample can be associated with a memory function $\mu(\tau)$ through a generalized Stokes-Einstein relation (GSER) that is given by~\cite{cordoba2012jrheol}
\begin{equation}
\langle \overline{\Delta r^2}(s) \rangle = \frac{2 d_e k_BT \overline{\mu}(s)}{s \left( s + \kappa  \overline{\mu}(s) \right)} ~~,
\label{GSER-MSD-mu}
\end{equation}
where $\langle \overline{\Delta r^2}(s) \rangle= \mathcal{L}[\langle \Delta x^2(\tau) \rangle;s]$ and $\overline{\mu}(s) = \mathcal{L}[\mu(\tau);s]$ are the Laplace transforms of the MSD and the memory kernel, respectively.
	Accordingly, by considering the local MSD given by Eq.~\ref{MSD-NM-KVMH-local}, the above GSER with $d_e=1$ yields a memory function that can be also written in terms of a Prony series~\cite{cordoba2012jrheol}, that is,
\begin{equation}
\mu_{\xi}(\tau) = \mu_{0,\xi} \delta(\tau) - \sum_{j=1}^{N} c_{j,\xi} \exp(-\tau/\Lambda_{j,\xi})~~,
\label{memory-kernel-xi}
\end{equation}
with $\delta(\tau)$ being the Dirac's delta function.
	Here, as in Ref.~\cite{rizzi2020jrheol}, the non-Markovian dynamics of probe particles immersed in a mesoscopic region of the viscoelastic semisolid material can be generally described by a region-dependent overdamped GLE that can be written as
\begin{equation}
\frac{dx(t)}{dt} = -\kappa \int_{-\infty}^{t} \mu_{\xi}(t-t^{\prime}) x(t^{\prime}) dt^{\prime} + g_{\xi}(t) ~~,
\label{GLE-local}
\end{equation}
where the random velocity $g_{\xi}(t)$ is characterized by zero mean $\langle g_{\xi}(t) \rangle = 0$ and is related to the memory kernel $\mu_{\xi}(t)$ through a generalized and $\xi$-dependent fluctuation-dissipation relation that is given by
\begin{equation}
\langle g_{\xi}(t) g_{\xi}(t^{\prime}) \rangle_{\text{eq}} = k_BT \mu_{\xi}(t-t^{\prime})~~.
\end{equation}

\section{GLE-based simulations}
\label{GLEsimulations}

	By considering the numerical method to simulate overdamped GLEs of Ref.~\cite{cordoba2012jrheol}, we assume that the non-Markovian dynamics of a trapped probe particle described by Eq.~\ref{GLE-local} can be simulated through a scheme that uses $N+1$ coupled stochastic differential equations (SDEs).
	In practice, if the local memory kernel can be written in terms of a Prony series like in Eq.~\ref{memory-kernel-xi}, one has that the $m$-th trajectory, which corresponds to the simulation of a single probe particle in a particular mesoscopic region of the sample indexed by $\vec{R}_m$ with microrheological properties denoted by $\xi_m = \xi(\vec{R}_m)$, can be sampled as
\begin{widetext}
\begin{equation}
x_m(t+\Delta t)
=
x_m(t)
- \left( \kappa \mu_{0,m} x_m(t) + \sum_{j=1}^{N} \frac{c_{j,m} Q_{j,m}(t)}{\mu_{0,m}} \right) \Delta t
+ \sqrt{ 2k_BT \left( \mu_{0,m} - \sum_{j=1}^{N} c_{j,m} \Lambda_{j,m} \right)} \, \epsilon_{0}(t)
+ \sum_{j=1}^{N} \sqrt{2k_BT c_{j,m} \Lambda_{j,m} } \, \epsilon_{j}(t) ~~,
\label{eq:x-update-scheme}
\end{equation}
with the auxiliary variables evaluated as
\begin{equation}
Q_{j,m}(t+\Delta t) = 
Q_{j,m}(t)
- \left( \kappa \mu_{0,m} x_m(t) + \frac{Q_{j,m}(t)}{\Lambda_{j,m}} \right) \Delta t + 
\sqrt{\frac{2 k_BT (\mu_{0,m})^2}{c_{j,m} \Lambda_{j,m}} } \, \epsilon_{j}(t)~~,
\label{eq:q-update-scheme} 
\end{equation}
\end{widetext}
where $\epsilon_{j}(t)=\text{N}_j(0,1)\sqrt{\Delta t}$ ($j=0,1,\dots,N$), with $\text{N}_j(0,1)$ being independent Gaussian random variables with zero mean and unitary variance.
	Initial conditions are taken as $x_m(0)=0$ and $Q_{j,m}(0)=0$, but in order to measure MSDs we consider trajectories at least twice as long as the longest relaxation time so that the first half of the time series can be discarded.

	As described in Appendix~\ref{mu-from-MSD}, if the MSD is written in terms of a Prony series as in Eq.~\ref{MSD-NM-KVMH-local}, one can obtain the parameters $\mu_{0,m}$, $\{ c_{j,m} \}$, and $\{ \Lambda_{j,m} \}$ (with $j=1,\dots,N$) used in the above equations from the parameters $\kappa$, $T$, $\{ q_{j,m} \}$ and $\{ \gamma_{j,m} \}$ (with $j=1,\dots,N+1$) used to define the MSD, as prescribed in Sec.~\ref{approxNM-KVMH}.
	It is worth noting that the parameters are not all independent of each other.
	For instance, since $\lim_{\tau \rightarrow 0} \langle \Delta x^2(\tau) \rangle_m=0$, one has the condition that 
\begin{equation}
\sum_{j=1}^{N+1} q_{j,m} = 2 \frac{k_BT}{\kappa}~~,
\label{q-condition-de1}
\end{equation}
for any mesoscopic region (see also Eq.~\ref{q-condition}).
	Besides, one should be sure that $\mu_{0,m} >\sum_{j=1}^{N} c_{j,m} \Lambda_{j,m} $, as required by Eq.~\ref{eq:x-update-scheme}.
	This condition is naturally satisfied if one follows the procedure described in Appendix~\ref{mu-from-MSD}, from where $\mu_{0,m}$ can be evaluated as in Eq.~\ref{mu0-formula} or, similarly, as
\begin{equation}
\mu_{0,m} = \frac{1}{2 k_BT} \sum_{j=1}^{N+1} \frac{q_{j,m}}{\gamma_{j,m}} ~~.
\label{mu0m-def}
\end{equation}
	Interestingly, the parameter $\mu_{0,m}$ can be thought of as a local mobility and can be used to define a diffusion coefficient as $D_m = \mu_{0,m} k_BT$.
	In a way, our approach revealed that the micro-heterogeneities in the viscoelastic semisolid material can be interpreted in terms of different local diffusion coefficients $D_m$, i.e., different local viscosities $\eta_{0,m}$.

\section{Numerical results}
\label{results}

%%%%%%%%%%%%%%%%%%%%%%%%%%%%%%%%%%%%%%%%%
%%%%%%%%%%%% FIG. 2: MSD Dtau %%%%%%%%%%%
%%%%%%%%%%%%%%%%%%%%%%%%%%%%%%%%%%%%%%%%%
\begin{figure}[!t]
\centering
\includegraphics[width=0.35\textwidth]{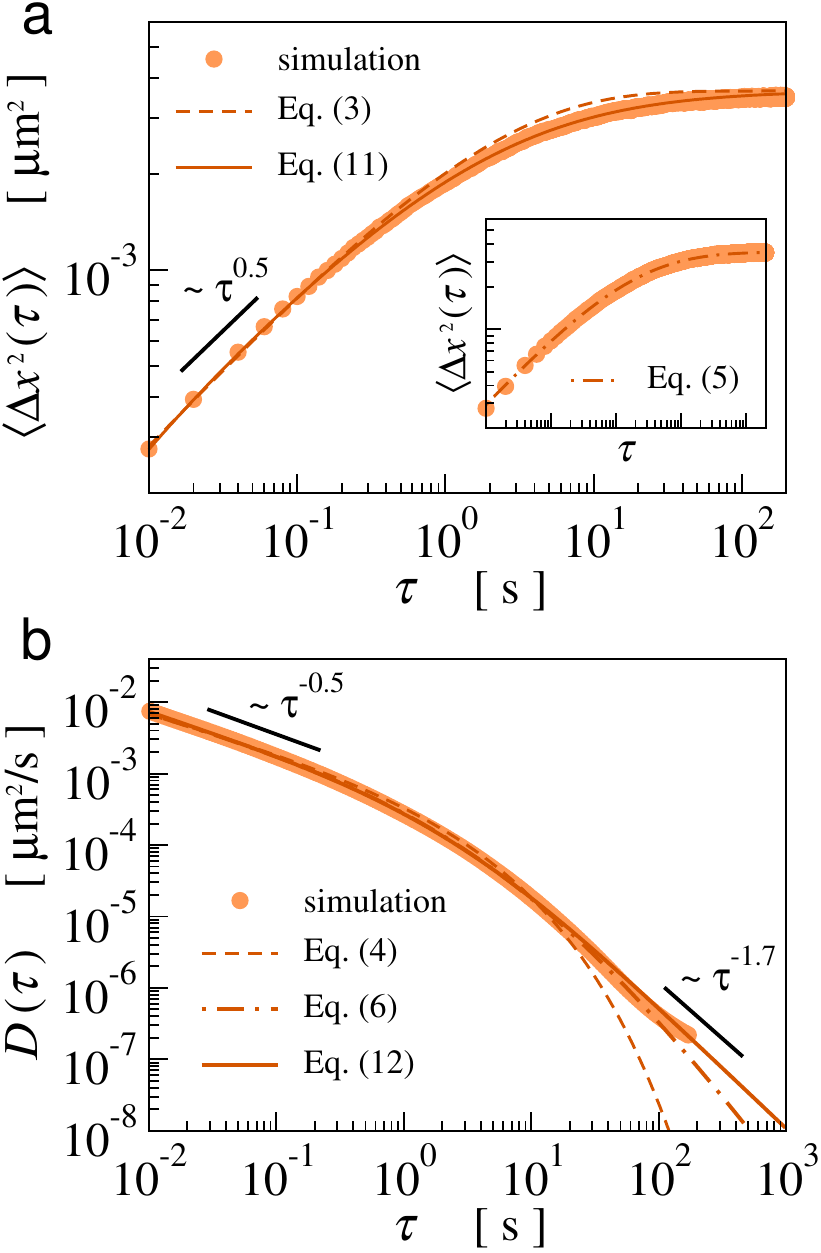}
\caption{(a)~Mean squared displacement $\langle \Delta x^2(\tau) \rangle$ and (b)~time-dependent diffusion coefficient $D(\tau)$.
    Filled circles correspond to numerical results obtained from GLE-based simulations, i.e., Eqs.~\ref{eq:x-update-scheme} and~\ref{eq:q-update-scheme}, and correspond to averages over $N_t=10^5$ independent trajectories.
	Continuous lines correspond to the theoretical results expected from the approximated NM-KVMH model, that is, Eqs.~\ref{MSD-NM-KVMH-final} and~\ref{Dtau-NM-KVMH-final} with $\kappa=2.261\,$pN/$\mu$m, $T=298\,$K, $n=0.5$, $p=0.7$, and the reference parameters  $q_j^{*}$ and $\gamma_j^{*}$ given in Table~\ref{tabela-referencia}.
	Dashed lines denote fits to the NM-KV model, Eqs.~\ref{MSD-NM-KV}, while dot-dashed line in the inset panel in (a) correspond to the fit of the NM-KVMH model, Eqs.~\ref{MSDjrheol2020}, both with the same values of $n$, $\kappa$, $T$, and $\tau_c$.
}
\label{fig:MSD-Dtau}
\end{figure}
%%%%%%%%%%%%%%%%%%%%%%%%%%%%%%%%%%%%%%%%%
%%%%%%%%%%%%%%%%%%%%%%%%%%%%%%%%%%%%%%%%%
%%%%%%%%%%%%%%%%%%%%%%%%%%%%%%%%%%%%%%%%%

	To validate our numerical approach, we consider probe particles with radius $a=3\,\mu$m in a semisolid material characterized by an effective spring constant equal to $\kappa=2.261\,$pN/$\mu$m at a temperature $T=298\,$K.
	We note that the parameters $q_j^{*}$ in Table~\ref{tabela-referencia} are taken to be consistent with these values of $T$ and $\kappa$ and an exponent equal to $n=0.5$, as expected from Eq.~\ref{q-condition-de1}.
	Equations~\ref{eq:x-update-scheme} and~\ref{eq:q-update-scheme} were used with $\Delta t=2 \times 10^{-5}\,$s to produce $N_t=10^{5}$ independent trajectories, each with a different set of parameters $q_{j,m}=q_j^{*}$ and $\gamma_{j,m}=(p/\xi_m)\gamma_j^{*}$.
	The values of the parameters $\mu_{0,m}$, $\{ c_{j,m} \}$, and $\{ \Lambda_{j,m}\}$ were re-evaluated for each independent trajectory through the scheme detailed in Appendix~\ref{mu-from-MSD}.
	The random variable $\xi_m$ was obtained from a generalized gamma distribution (Eq.~\ref{distribution-rho-xi}) with $p=0.7$, which characterizes the micro-heterogeneities within the sample, so that the final MSD corresponds to an average over all trajectories, emulating a real microrheology experiment.

%%%%%%%%%%%%%%%%%%%%%%%%%%%%%%%%%%%%%%%%%
%%%%%%%%%%%% FIG. 3: dist %%%%%%%%%%%
%%%%%%%%%%%%%%%%%%%%%%%%%%%%%%%%%%%%%%%%%
\begin{figure*}[!t]
\centering
\includegraphics[width=0.75\textwidth]{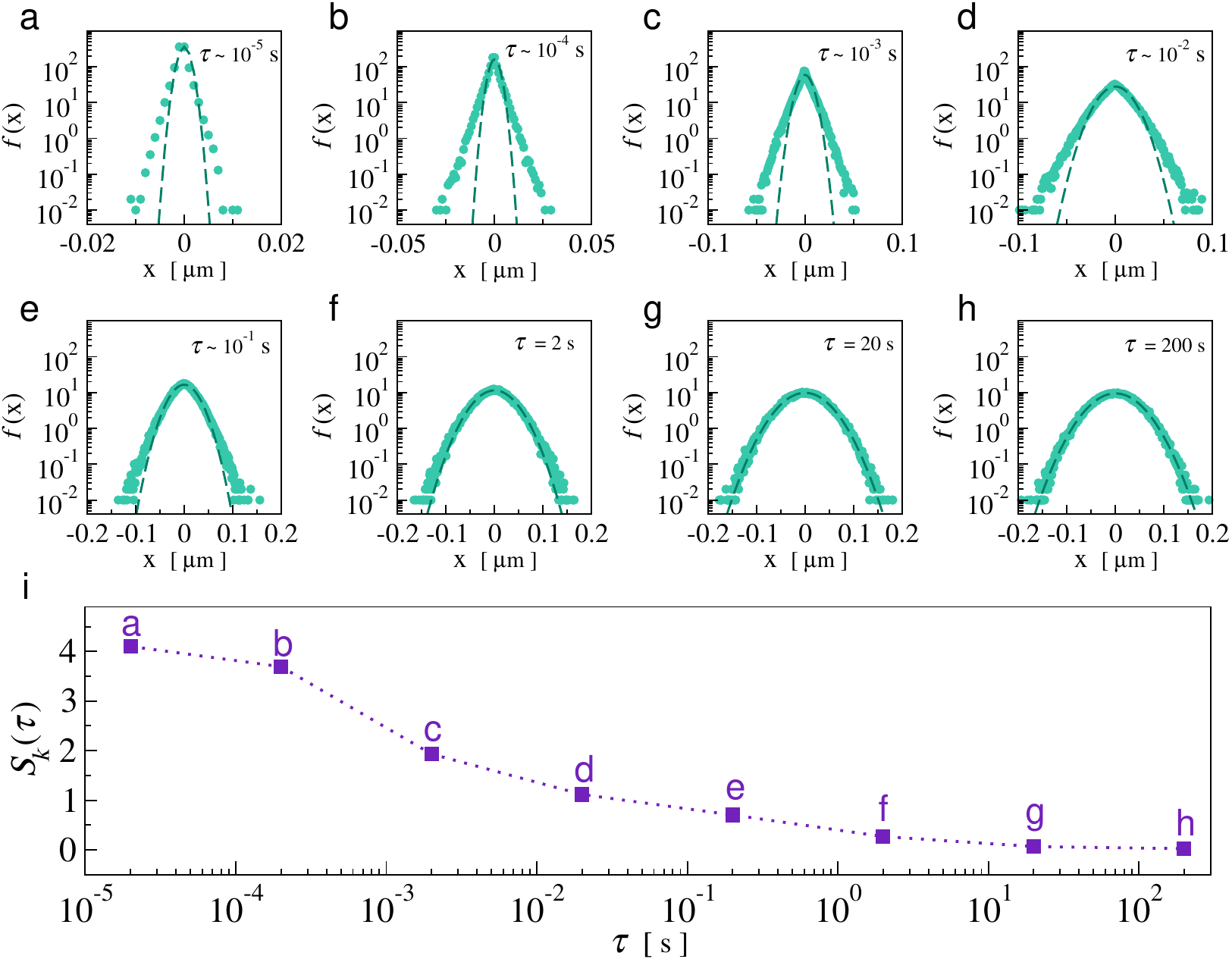}
\caption{(a)-(h)~Displacement (i.e., van Hove) distributions $f(x,\tau)$ at different times $\tau$ obtained numerically from the GLE-based simulations.
    Filled circles denote histograms computed from $N_t=10^5$ trajectories with $\kappa=2.261\,$pN/$\mu$m, $T=298\,$K, $\Delta t=2 \times 10^{-5}\,$s, $p=0.7$, and the reference parameters $q_j^{*}$ and $\gamma_j^{*}$ given in Table~\ref{tabela-referencia}.
    Dashed lines denote fitted Gaussian distributions.
    (i) Excess of kurtosis $S_k(\tau)$, Eq.~\ref{excesskurt}.
}
\label{fig:distributions}
\end{figure*}
%%%%%%%%%%%%%%%%%%%%%%%%%%%%%%%%%%%%%%%%%
%%%%%%%%%%%%%%%%%%%%%%%%%%%%%%%%%%%%%%%%%
%%%%%%%%%%%%%%%%%%%%%%%%%%%%%%%%%%%%%%%%%

	Figures~\ref{fig:MSD-Dtau}(a) and~(b) display results obtained from our GLE-based simulations for both the MSD $\langle \Delta x^2(\tau) \rangle$ and the time-dependent diffusion coefficient $D(\tau)$, where
the latter was computed numerically as $D(\tau)=\langle \Delta x^2(\tau) \rangle p(\tau)/2\tau$,
with $p(\tau) = d \ln \langle \Delta x^2(\tau) \rangle/d \ln \tau$ (as the experimental ones presented in Fig.~\ref{examples}).
    The agreement between the numerical data to the expressions derived here for the approximated NM-KVMH model, i.e., Eqs.~\ref{MSD-NM-KVMH-final} and~\ref{Dtau-NM-KVMH-final}, validates the non-Markovian numerical scheme used to study the influence of the micro-heterogeneities and non-Markovianity in the microrheology of an effective semisolid material.
	The comparison between the numerical results and the fitted expressions obtained for the NM-KV is qualitatively similar to what is illustrated with experimental data in Fig.~\ref{examples} and indicates that Eqs.~\ref{MSD-NM-KV} and~\ref{difcoefNM-KV} do not describe the numerical data very well when micro-heterogeneities are present in the sample.
	Conversely, the inset panel in Fig.~\ref{fig:MSD-Dtau}(a) shows that Eq.~\ref{MSDjrheol2020} can be used to describe the numerical data obtained for the MSD very well.
	Even so, the exponent $\alpha\approx 3$ obtained from this fit led to a slightly different power-law behavior of $D(\tau)$ at later times.
	This is because the Prony series is an approximation introduced to perform the simulations and to derive Eqs.~\ref{MSD-NM-KVMH-final} and~\ref{Dtau-NM-KVMH-final}.
	Indeed, Fig.~\ref{fig:MSD-Dtau}(b) shows that, while Eq.~\ref{Dtaurheol2020} predicts $D(\tau) \propto \tau^{-2.5}$, Eq.~\ref{Dtau-NM-KVMH-final} leads to $D(\tau) \propto \tau^{-0.7}$, and the numerical results in Figure~\ref{fig:MSD-Dtau}(b) seem to change from one regime to the other at later times.

    Next, to further analyze the influence of the micro-heterogeneities in the microrheology of such an effective semisolid material, we include in panels (a)-(h) of Fig.~\ref{fig:distributions} the displacement (i.e., van Hove) distributions $f(x,\tau)$ obtained from our simulations.
	To assess the deviation from Gaussian distributions, we also include in Fig.~\ref{fig:distributions} the corresponding fitted Gaussian curves (dashed lines) and the excess of kurtosis~\cite{donald2008eurphysJE}
\begin{equation}
    S_k(\tau) = \sum_{j=1}^{N_t} \frac{[x_j(\tau)-\bar{x}(\tau)]^4}{(N_t-1)[\sigma_x(\tau)]^4}-3 ~~,
    \label{excesskurt}
\end{equation}
where $\bar{x}(\tau)$ is the mean and $\sigma_x(\tau)$ is the standard deviation evaluated from all the $N_t$ trajectories at a time $\tau$.
	From $S_k(\tau)$ one has that the higher the excess of kurtosis the larger the deviation from a Gaussian distribution.
    	And, as is shown in Fig.~\ref{fig:distributions}(i), $S_k(\tau)$ is higher at short times and 
decreases for later times, indicating that, in our simulations, the micro-heterogeneities strongly affect the short-time displacement distributions, in agreement with what is observed from microrheology experiments on semisolid materials~\cite{donald2009eurphysJE,donald2008eurphysJE}.
    
\begin{figure}[!t]
\centering
\includegraphics[width=0.47\textwidth]{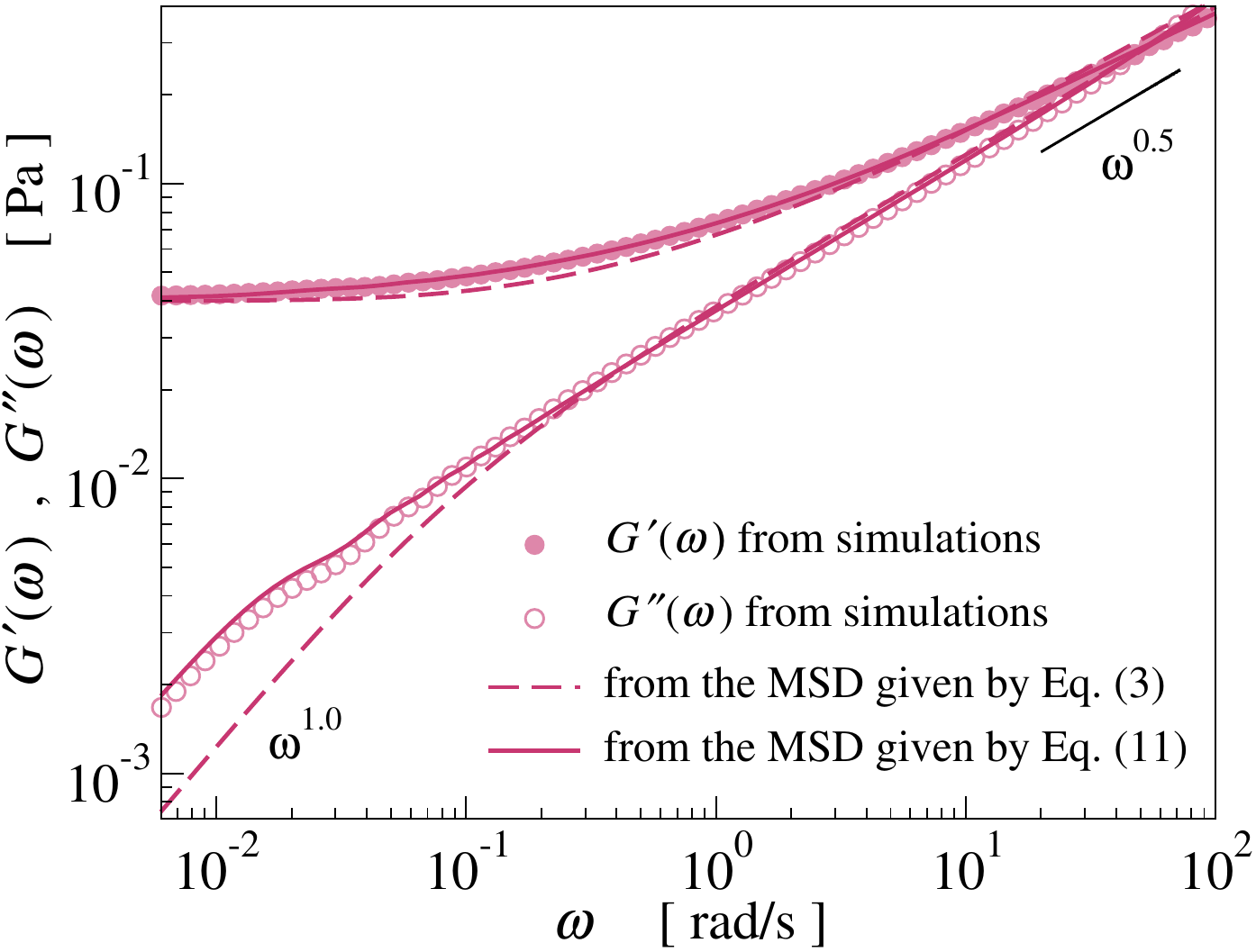}
\caption{
        Filled and empty circles correspond to the storage modulus $G^{\prime}(\omega)$ and the loss modulus $G^{\prime \prime}(\omega)$ obtained from Fourier transform of the compliance $J(\tau)$ obtained from the MSD $\langle \Delta x^2 (\tau) \rangle$, Eq.~\ref{GSER-Jtau-MSD}, through the numerical method of Ref.~\cite{Evans2009} (see also Ref.~\cite{azevedo2025softmatter}). 
        Dashed lines denote estimates obtained from the NM-KV model, Eq.~\ref{MSD-NM-KV}, while continuous lines were determined from the approximated NM-KVMH, Eq.~\ref{MSD-NM-KVMH-final}, both with the parameters used to obtain the results shown in Fig.~\ref{fig:MSD-Dtau}.
}
\label{fig:shearmoduli}
\end{figure}

	Finally, we include in Fig.~\ref{fig:shearmoduli} the estimates for the storage modulus, $G^{\prime}(\omega)$, and for the loss modulus, $G^{\prime \prime}(\omega)$, which were evaluated numerically using the method described in Ref.~\cite{Evans2009} (see also Ref.~\cite{azevedo2025softmatter} for further details).
	The numerical data and the different curves were obtained from the GSER given by Eq.~\ref{GSER-Jtau-MSD}, which links the MSD to the compliance of the material $J(\tau)$.
	As expected, the results in Fig.~\ref{fig:shearmoduli} indicate that both the MSD obtained from the NM-KV model, Eq.~\ref{MSD-NM-KV}, and from the approximated NM-KVMH, Eq.~\ref{MSD-NM-KVMH-final}, lead to a semisolid rheological behavior in which the storage modulus displays a plateau regime, i.e., $G_0=\lim_{\omega \rightarrow 0} G^{\prime}(\omega)$, and a power-law regime at high frequencies, where $G^{\prime}(\omega) \propto \omega^n$.
	Just like the behavior illustrated by the data presented in Fig.~\ref{examples} for the storage modulus $G^{\prime}(\omega)$, one has that the micro-heterogeneities smooth the transition between these two regimes.
	In addition, the power-law behavior observed for the loss modulus at high frequencies, i.e., $G^{\prime \prime}(\omega) \propto \omega^n$, is similar to the microrheology obtained from the original NM-KVMH model (see Fig.~\ref{examples}).
	The main difference between the models with and without micro-heterogeneities is observed for $G^{\prime \prime}(\omega)$ at low frequencies.
	As mentioned in Sec.~\ref{GLEsimulations}, one can interpret the micro-heterogeneities introduced in the approximated NM-KVMH model in terms of different local viscosities (i.e., different local diffusion coefficients), and our results revealed that their effect can be clearly observable in $G^{\prime \prime}(\omega)$, perhaps because the loss modulus is related to the frequency-dependent viscosity~\cite{ferrybook} as $G^{\prime \prime}(\omega)=\eta^{\prime}(\omega)/\omega$.
	Figure~\ref{fig:shearmoduli} indicates that the results for the loss modulus obtained from both the approximated NM-KVMH and the NM-KV exhibit a Maxwell-like behavior where $\lim_{\omega \rightarrow 0}\eta^{\prime}(\omega) \approx \eta_{0,\text{eff}}$, even though with different effective viscosities, $\eta_{0,\text{eff}}^{\text{NM-KVMH}} > \eta_{0,\text{eff}}^{\text{NM-KV}}$.
	Interestingly, by considering the MSD of the original NM-KVMH~\cite{rizzi2020jrheol}, Eq.~\ref{MSDjrheol2020}, with the same parameters of the NM-KV and an exponent $\alpha \approx 3$ so that $\alpha n \approx 1.5$, one also gets a linear behavior for $G^{\prime \prime}(\omega)$ (data not shown) instead of the power-law behavior $G^{\prime \prime}(\omega) \propto \omega^{\alpha n}$ displayed when $\alpha n < 1$ (see, e.g., the loss moduli in Fig.~\ref{examples}).

%%%%%%%%%%%%%%%%%%%%%%%%%%%%%%%%%%%%%%%%%
%%%%%%%%%%%% FIG. 4: shear moduli %%%%%%%%%%%
%%%%%%%%%%%%%%%%%%%%%%%%%%%%%%%%%%%%%%%%%
%%%%%%%%%%%%%%%%%%%%%%%%%%%%%%%%%%%%%%%%%
%%%%%%%%%%%%%%%%%%%%%%%%%%%%%%%%%%%%%%%%%
%%%%%%%%%%%%%%%%%%%%%%%%%%%%%%%%%%%%%%%%%

\section{Concluding remarks}
\label{conclusions}

	As illustrated by the experimental data presented in Fig.~\ref{examples}, the microrheology of isotropic semisolid viscoelastic materials is characterized by the power-law behavior of the MSD at short times, i.e., $\langle \Delta r^2(\tau) \rangle \propto \tau^n$, which leads to the power-law rheology, i.e., $G^{\prime} \propto G^{\prime \prime} \propto \omega^n$, at high frequencies.
	Based on these and other experimental evidences~\cite{jahnel2008softmatter,georgiades2013biopolymers}, we assume that the sub-diffusive behavior displayed by the MSD of probe particles immersed in semisolids corresponds to the hallmark of non-Markovianity in the dynamics of the structures in the material, and, as one can infer from the numerical results of Ref.~\cite{duarte2021sm}, the flexibility of the structures may lead to different exponents $n$ ranging from $0.5$ and $0.75$.
	Interestingly, by considering a memory kernel given by a Prony series, we were able to perform non-Markovian Brownian simulations through a GLE-based scheme that emulates the microrheology probed by particles in contact with such structures, since Eq.~\ref{memory-kernel-xi} leads to a relaxation modulus $G(\tau)$ that is also written in terms of a Prony series (see Ref.~\cite{cordoba2012jrheol} for details).
	Incidentally, the non-Markovian dynamics of flexible chains such as those described by the Rouse model~\cite{rubinsteinbook} can be directly associated with a relaxation modulus $G(\tau)$ that is given by a sum of exponentials and displays a power-law decay at intermediate times with exponent $n=0.5$ which leads to shear moduli that are proportional to $\omega^{0.5}$.
	Hence, our non-Markovian interpretation of the microrheology of semisolid viscoelastic materials has a physically motivated framework which is distinct from other approaches to sub-diffusion and power-law rheology, e.g., caging in porous/fractal substrates~\cite{BenAvraham2002} and effective fractional models~\cite{bonfanti2020softmatter}.

	At last, it is worth mentioning that, although non-Gaussian displacement distributions have been experimentally observed in the microrheology of semisolid viscoelastic materials and interpreted in terms of heterogeneities~\cite{donald2008eurphysJE,donald2009eurphysJE,donald2012softmatter,donald2014eurphysJE}, a more formal theoretical framework has only recently been developed.
	From the ideas of our previous studies~\cite{rizzi2020jrheol,azevedo2025softmatter} and the analysis of the experimental data presented in Fig.~\ref{examples}, we established that the hallmark of the influence of micro-heterogeneities on the dynamics of probe particles immersed in semisolids is a smooth crossover between the power-law and plateau regimes in the mean squared displacement $\langle \Delta r^2(\tau) \rangle$, which leads to the more identifiable power-law behavior observed in the time-dependent diffusion coefficient $D(\tau)$ at later times.
	By considering the GLE-based scheme implemented here, we found a reliable way to sample single trajectories of probe particles trapped at mesoscopic regions with different microrheological properties, just as in experimental microrheology.
	Hence, the simulations allowed us to explore how the micro-heterogeneities affect not only the average values of the MSD and $D(\tau)$ but also the displacement distributions $f(x,\tau)$.
	Accordingly, our numerical results validate both the presence of non-Gaussian distributions and the non-exponential behavior of the time-dependent diffusion coefficient at later times.
	We note that, although we have considered an approximated version of the NM-KVMH in terms of a Prony series to perform and validate the numerical simulations, it may be more practical to fit the experimental data to the expressions of the original NM-KVMH, as we did in Fig.~\ref{examples}.
	Besides providing a convenient way to determine the degree of heterogeneity of the sample by a single parameter, i.e., the exponent $\alpha$, the analytical expressions of the NM-KVMH, i.e., Eqs.~\ref{MSDjrheol2020} and~\ref{Dtaurheol2020}, can also be used as input to Eq.~\ref{GSER-Jtau-MSD} to obtain reliable estimates for the compliance and shear moduli of the semisolids when the experimental data are noisy.
    In terms of mechanical response, our results indicate that the hallmark of the presence of micro-heterogeneities in semisolid viscoelastic materials corresponds to the power-law behavior observed in the loss modulus $G^{\prime \prime}$ at low frequencies. 
    In contrast to the NM-KV model and other models such as the fractional KV and ladder models~\cite{delgadonatphys}, where the loss modulus displays a linear behavior, i.e., $G^{\prime \prime} \propto \omega$, the analysis of the experimental data in Fig.~\ref{examples} and the numerical results in Ref.~\cite{delgadonatphys} indicate that in many cases the power-law behavior described by the NM-KVMH model~\cite{rizzi2020jrheol}, i.e.,  $G^{\prime \prime} \propto \omega^{\alpha n}$, may describe the microrheology of heterogeneous semisolids more accurately.

\begin{acknowledgments}
T. N. Azevedo thanks CAPES for the studentship and L. G. Rizzi acknowledges the support from CNPq (Grants No. 312999/2021-6 and 308285/2025-5) and FAPEMIG (Process APQ-01462-24). The authors also thank the computational resources made available by GISC-UFV.
\end{acknowledgments}

%\section*{Data Availability Statement}

%The data that support the findings of this study are available from the corresponding author upon reasonable request.

\appendix

\section{Memory function from the MSD}
\label{mu-from-MSD}

	In this Appendix we show how one can obtain the parameters $\mu_{0,m}$, $\{ c_{j,m} \}$, and $\{ \Lambda_{j,m} \}$ of the memory function from the parameters $\kappa$, $k_BT$, $\{ q_{j,m} \}$, and $\{ \gamma_{j,m} \}$ of the local MSD to perform the GLE-based simulations described in Sec.~\ref{GLEsimulations}.

	The idea behind Eqs.~\ref{eq:x-update-scheme} and~\ref{eq:q-update-scheme} is that the dynamics of a probe particle in the $m$-trajectory is governed by a memory kernel like in Eq.~\ref{memory-kernel-xi} and can be written through a Prony series as $\mu_{\xi_m}(\tau) = \mu_{0,m} \delta(\tau) - \sum_{j=1}^{N} c_{j,m} \exp(-\tau/\Lambda_{j,m})$, where the parameters are associated to a specific mesoscopic region of the sample with microrheological properties indexed by $\xi_m = \xi(\vec{R}_m)$.
	Following Ref.~\cite{cordoba2012jrheol}, one finds that its Laplace transform can be written as
\begin{equation}
 \overline{\mu}(s)  = \mu_{0,m} - \frac{P(s)}{Q(s)}~~
\end{equation}
where $P(s)$ and $Q(s)$ are polynomials given by
\begin{equation}
P(s) = \sum_{i=1}^{N} c_{i,m} \prod_{j \neq i}^{N}(s + 1/\Lambda_{j,m})~~,
\label{Ps-def}
\end{equation}
and
\begin{equation}
Q(s) = \prod_{j = 1}^{N}(s + 1/\Lambda_{j,m})~~.
\label{Qs-def}
\end{equation}
	Similarly, we assume that the MSD is given as in Eq.~\ref{MSD-NM-KVMH-local}, that is, $\langle \Delta r^2(\tau) \rangle = 2 \left(d_e k_B T/\kappa \right) - \sum_{j=1}^{N+1} q_{j,m} \exp(-\tau/\gamma_{j,m})$, where the parameters of the $m$-trajectory are given as in Sec.~\ref{approxNM-KVMH}, that is, $q_{j,m} = q_{j}^{*}$ and $\gamma_{j,m} = (p/\xi_m) \gamma_{j}^{*}$ (as in Eq.~\ref{local-gamma-xi}), with $\{ q_{j}^{*} \}$ and $\{ \gamma_{j}^{*} \}$ given from Table~\ref{tabela-referencia}, and $\xi_m$ being a random variable obtained according to the generalized gamma distribution, Eq.~\ref{distribution-rho-xi}.
	Hence, one has that its Laplace transform can be written as
\begin{equation}
\langle \overline{\Delta r^2}(s) \rangle = 
2 \frac{d_e k_BT }{\kappa} \frac{1}{s} - \frac{E(s)}{F(s)}~~,
\end{equation}
where $E(s)$ and $F(s)$ are polynomials given by
\begin{equation}
E(s) = \sum_{i=1}^{N+1} q_{i,m} \prod_{j \neq i}^{N+1}(s + 1/\gamma_{j,m})~~,
\label{Es-def}
\end{equation}
and
\begin{equation}
F(s) = \prod_{j = i}^{N+1}(s + 1/\gamma_{j,m})~~.
\label{Fs-def}
\end{equation}

	Now, by considering the GSER given by Eq.~\ref{GSER-MSD-mu} that relates the Laplace transforms of the MSD $\langle \overline{\Delta r^2}(s) \rangle$ and the memory function $\overline{\mu}(s)$, one can write the polynomials of the memory function in terms of the polynomials of the MSD, that is,
\begin{equation}
P(s) = \left( \frac{\mu_{0,m} \kappa + s}{k_BT} \right) E(s) - \frac{2 d_e}{\kappa} F(s) ~~,
\label{Ps-Es-Fs}
\end{equation}
and
\begin{equation}
Q(s) = \frac{\kappa}{k_BT} E(s) ~~.
\label{Qs-Es}
\end{equation}

%{\color{red} explicar que 1.}
%	From the condition that
%% /home/lerizzi/Documents/Producao_2024/TIAGO_DOC/20240929/to_tiago/parametros.pdf
%\begin{equation}
%\sum_{j=1}^{N+1} q_j = 2 \frac{d_e k_BT}{\kappa}~~,
%\end{equation}
%{\color{red} e que 2.}
%and
%\begin{equation}
%\mu_0 = \frac{1}{2 k_BT} \sum_{j=1}^{N+1} \frac{q_j}{\gamma_j}~~.
%\end{equation}
%or, equivalently, {\color{red} com base na equivalencia dos graus dos polinomios}
%% /home/lerizzi/Documents/Producao_2024/TIAGO_DOC/20241011/gle_simulation_cordoba2012_7modes.pdf
	Interestingly, because $P(s)$ should be a  $(N-1)$-degree polynomial (see Eq.~\ref{Ps-def}) and the right-hand side of Eq.~\ref{Ps-Es-Fs} is a $(N+1)$-degree polynomial, the identity leads to the following relations:
\begin{equation}
\sum_{j=1} q_{j,m} = 2 \frac{d_e k_BT}{\kappa}~~,
\label{q-condition}
\end{equation}
and
\begin{equation}
\mu_{0,m} = \frac{1}{\kappa} \sum_{j=1}^{N+1} \frac{1}{\gamma_{j,m}} - \frac{1}{2 k_BT} \sum_{j=1}^{N+1} q_{j,m} \sum_{k \neq j}^{N+1} \frac{1}{\gamma_{j,m}}~~,
\label{mu0-formula}
\end{equation}
with the latter being equivalent to Eq.~\ref{mu0m-def}.

	The next step is to replace the definitions of $Q(s)$, Eq.~\ref{Qs-def}, and $E(s)$, Eq.~\ref{Es-def}, in Eq.~\ref{Qs-Es}, which yields the following identity
\begin{equation}
Q(s) = \prod_{j = 1}^{N}(s + 1/\Lambda_{j,m}) = \frac{\kappa}{k_BT} \sum_{i=1}^{N+1} q_{i,m} \prod_{j \neq i}^{N+1}(s + 1/\gamma_{j,m})~~.
\end{equation}
	Thus, the $N$ roots of the polynomial at the right-hand side of the above equation, which are given in terms of $\kappa$, $k_BT$, $\{ q_{j,m} \}$, and $\{ \gamma_{j,m} \}$, should provide the values of the characteristic times $\Lambda_{j,m}$, which correspond to the roots of the polynomial at the left-hand side.

	In addition, since the degree of $Q(s)$ is greater than $P(s)$, their ratio can be written through partial fraction decomposition as
\begin{equation}
\frac{P(s)}{Q(s)} = \sum_{j=1}^{N} \frac{P(\alpha_j)}{Q^{\prime}(\alpha_j)} \frac{1}{s-\alpha_j} ~~,
\end{equation}
where $\alpha_j=-1/\Lambda_{j,m}$ and the prime indicates the derivative $Q^{\prime}(s)=dQ(s)/ds$.
	Hence, with the $N$ values of $\Lambda_{j,m}$, one can compute the coefficients $c_{j,m} = P(-1/\Lambda_{j,m})/Q^{\prime}(-1/\Lambda_{j,m})$ as
\begin{equation}
c_{j,m} = %\frac{P(-1/\Lambda_j)}{Q^{\prime}(-1/\Lambda_j)} = 
\frac{
\left( \mu_{0,m} \kappa -\frac{1}{\Lambda_{j,m}}  \right) E\left(-\frac{1}{\Lambda_{j,m}}\right) - (2d_ek_BT/\kappa) F\left(-\frac{1}{\Lambda_{j,m}} \right)
}{\kappa 
 E^{\prime}\left(-\frac{1}{\Lambda_{j,m}} \right)
}~~ ,
\end{equation}
where $E(s)$ and $F(s)$ are the known polynomials that can be computed from Eqs.~\ref{Es-def} and~\ref{Fs-def}, respectively.
	Numerically, it is advisable to check if the condition expected from Eq.~\ref{eq:x-update-scheme} was attained, that is, $\mu_{0,m} > \sum_{j=1}^{N}c_{j,m} \Lambda_{j,m}$.
    % advisable or prudent

\section*{References}

%\nocite{*}
%\bibliography{aipsamp}% Produces the bibliography via BibTeX.
%\bibliography{biblio}% Produces the bibliography via BibTeX.

\begin{thebibliography}{36}%
\makeatletter
\providecommand \@ifxundefined [1]{%
 \@ifx{#1\undefined}
}%
\providecommand \@ifnum [1]{%
 \ifnum #1\expandafter \@firstoftwo
 \else \expandafter \@secondoftwo
 \fi
}%
\providecommand \@ifx [1]{%
 \ifx #1\expandafter \@firstoftwo
 \else \expandafter \@secondoftwo
 \fi
}%
\providecommand \natexlab [1]{#1}%
\providecommand \enquote  [1]{``#1''}%
\providecommand \bibnamefont  [1]{#1}%
\providecommand \bibfnamefont [1]{#1}%
\providecommand \citenamefont [1]{#1}%
\providecommand \href@noop [0]{\@secondoftwo}%
\providecommand \href [0]{\begingroup \@sanitize@url \@href}%
\providecommand \@href[1]{\@@startlink{#1}\@@href}%
\providecommand \@@href[1]{\endgroup#1\@@endlink}%
\providecommand \@sanitize@url [0]{\catcode `\\12\catcode `\$12\catcode
  `\&12\catcode `\#12\catcode `\^12\catcode `\_12\catcode `\%12\relax}%
\providecommand \@@startlink[1]{}%
\providecommand \@@endlink[0]{}%
\providecommand \url  [0]{\begingroup\@sanitize@url \@url }%
\providecommand \@url [1]{\endgroup\@href {#1}{\urlprefix }}%
\providecommand \urlprefix  [0]{URL }%
\providecommand \Eprint [0]{\href }%
\providecommand \doibase [0]{http://dx.doi.org/}%
\providecommand \selectlanguage [0]{\@gobble}%
\providecommand \bibinfo  [0]{\@secondoftwo}%
\providecommand \bibfield  [0]{\@secondoftwo}%
\providecommand \translation [1]{[#1]}%
\providecommand \BibitemOpen [0]{}%
\providecommand \bibitemStop [0]{}%
\providecommand \bibitemNoStop [0]{.\EOS\space}%
\providecommand \EOS [0]{\spacefactor3000\relax}%
\providecommand \BibitemShut  [1]{\csname bibitem#1\endcsname}%
\let\auto@bib@innerbib\@empty
%</preamble>
\bibitem [{\citenamefont {Ferry}(1980)}]{ferrybook}%
  \BibitemOpen
  \bibfield  {author} {\bibinfo {author} {\bibfnamefont {J.~D.}\ \bibnamefont
  {Ferry}},\ }\href@noop {} {\emph {\bibinfo {title} {Viscoelastic Properties
  of Polymers}}},\ \bibinfo {edition} {3rd}\ ed.\ (\bibinfo  {publisher} {John
  Wiley \& Sons, New York},\ \bibinfo {year} {1980})\BibitemShut {NoStop}%
\bibitem [{\citenamefont {Rao}(2014)}]{raobook}%
  \BibitemOpen
  \bibfield  {author} {\bibinfo {author} {\bibfnamefont {M.~A.}\ \bibnamefont
  {Rao}},\ }\href@noop {} {\emph {\bibinfo {title} {Rheology of Fluid,
  Semisolid, and Solid Foods: Principles and Applications}}},\ \bibinfo
  {edition} {3rd}\ ed.\ (\bibinfo  {publisher} {Springer, New York},\ \bibinfo
  {year} {2014})\BibitemShut {NoStop}%
\bibitem [{\citenamefont {Rizzi}\ and\ \citenamefont
  {Tassieri}(2018)}]{Rizzi_Tassieri_2018}%
  \BibitemOpen
  \bibfield  {author} {\bibinfo {author} {\bibfnamefont {L.~G.}\ \bibnamefont
  {Rizzi}}\ and\ \bibinfo {author} {\bibfnamefont {M.}~\bibnamefont
  {Tassieri}},\ }\bibfield  {title} {\enquote {\bibinfo {title} {Microrheology
  of biological specimens},}\ }in\ \href@noop {} {\emph {\bibinfo {booktitle}
  {Encyclopedia of Analytical Chemistry}}}\ (\bibinfo  {publisher} {John Wiley
  \& Sons},\ \bibinfo {year} {2018})\ pp.\ \bibinfo {pages} {1--24}\BibitemShut
  {NoStop}%
\bibitem [{\citenamefont {Azevedo}\ \emph {et~al.}(2025)\citenamefont
  {Azevedo}, \citenamefont {Oliveira}, \citenamefont {Maia}, \citenamefont
  {Teixeira},\ and\ \citenamefont {Rizzi}}]{azevedo2025softmatter}%
  \BibitemOpen
  \bibfield  {author} {\bibinfo {author} {\bibfnamefont {T.~N.}\ \bibnamefont
  {Azevedo}}, \bibinfo {author} {\bibfnamefont {K.~M.}\ \bibnamefont
  {Oliveira}}, \bibinfo {author} {\bibfnamefont {H.~P.}\ \bibnamefont {Maia}},
  \bibinfo {author} {\bibfnamefont {A.~V. N.~C.}\ \bibnamefont {Teixeira}}, \
  and\ \bibinfo {author} {\bibfnamefont {L.~G.}\ \bibnamefont {Rizzi}},\
  }\bibfield  {title} {\enquote {\bibinfo {title} {Microrheological model for
  \uppercase{K}elvin-\uppercase{V}oigt materials with micro-heterogeneities},}\
  }\href@noop {} {\bibfield  {journal} {\bibinfo  {journal} {Soft Matter}\
  }\textbf {\bibinfo {volume} {21}},\ \bibinfo {pages} {1498} (\bibinfo {year}
  {2025})}\BibitemShut {NoStop}%
\bibitem [{\citenamefont {Azevedo}\ and\ \citenamefont
  {Rizzi}(2020)}]{azevedo2020jphysconfser}%
  \BibitemOpen
  \bibfield  {author} {\bibinfo {author} {\bibfnamefont {T.~N.}\ \bibnamefont
  {Azevedo}}\ and\ \bibinfo {author} {\bibfnamefont {L.~G.}\ \bibnamefont
  {Rizzi}},\ }\bibfield  {title} {\enquote {\bibinfo {title} {Microrheology of
  filament networks from \uppercase{B}rownian dynamics simulations},}\
  }\href@noop {} {\bibfield  {journal} {\bibinfo  {journal} {J. Phys.: Conf.
  Ser.}\ }\textbf {\bibinfo {volume} {1483}},\ \bibinfo {pages} {012001}
  (\bibinfo {year} {2020})}\BibitemShut {NoStop}%
\bibitem [{\citenamefont {Rizzi}(2020)}]{rizzi2020jrheol}%
  \BibitemOpen
  \bibfield  {author} {\bibinfo {author} {\bibfnamefont {L.~G.}\ \bibnamefont
  {Rizzi}},\ }\bibfield  {title} {\enquote {\bibinfo {title} {Microrheological
  approach for the viscoelastic response of gels},}\ }\href@noop {} {\bibfield
  {journal} {\bibinfo  {journal} {J. Rheol.}\ }\textbf {\bibinfo {volume}
  {64}},\ \bibinfo {pages} {969} (\bibinfo {year} {2020})}\BibitemShut
  {NoStop}%
\bibitem [{\citenamefont {Jahnel}, \citenamefont {Waigh},\ and\ \citenamefont
  {Lu}(2008)}]{jahnel2008softmatter}%
  \BibitemOpen
  \bibfield  {author} {\bibinfo {author} {\bibfnamefont {M.}~\bibnamefont
  {Jahnel}}, \bibinfo {author} {\bibfnamefont {T.~A.}\ \bibnamefont {Waigh}}, \
  and\ \bibinfo {author} {\bibfnamefont {J.}~\bibnamefont {Lu}},\ }\bibfield
  {title} {\enquote {\bibinfo {title} {Thermal fluctuations of fibrin fibres at
  short time scales},}\ }\href@noop {} {\bibfield  {journal} {\bibinfo
  {journal} {Soft Matter}\ }\textbf {\bibinfo {volume} {4}},\ \bibinfo {pages}
  {1438--1442} (\bibinfo {year} {2008})}\BibitemShut {NoStop}%
\bibitem [{\citenamefont {Georgiades}\ \emph {et~al.}(2013)\citenamefont
  {Georgiades}, \citenamefont {Pudney}, \citenamefont {Thornton},\ and\
  \citenamefont {Waigh}}]{georgiades2013biopolymers}%
  \BibitemOpen
  \bibfield  {author} {\bibinfo {author} {\bibfnamefont {P.}~\bibnamefont
  {Georgiades}}, \bibinfo {author} {\bibfnamefont {P.~D.~A.}\ \bibnamefont
  {Pudney}}, \bibinfo {author} {\bibfnamefont {D.~J.}\ \bibnamefont
  {Thornton}}, \ and\ \bibinfo {author} {\bibfnamefont {T.~A.}\ \bibnamefont
  {Waigh}},\ }\bibfield  {title} {\enquote {\bibinfo {title} {Particle tracking
  microrheology of purified gastrointestinal mucins},}\ }\href@noop {}
  {\bibfield  {journal} {\bibinfo  {journal} {Biopolymers}\ }\textbf {\bibinfo
  {volume} {101}},\ \bibinfo {pages} {366} (\bibinfo {year}
  {2013})}\BibitemShut {NoStop}%
\bibitem [{\citenamefont {Duarte}, \citenamefont {Teixeira},\ and\
  \citenamefont {Rizzi}(2021)}]{duarte2021sm}%
  \BibitemOpen
  \bibfield  {author} {\bibinfo {author} {\bibfnamefont {L.~K.~R.}\
  \bibnamefont {Duarte}}, \bibinfo {author} {\bibfnamefont {A.~V. N.~C.}\
  \bibnamefont {Teixeira}}, \ and\ \bibinfo {author} {\bibfnamefont {L.~G.}\
  \bibnamefont {Rizzi}},\ }\bibfield  {title} {\enquote {\bibinfo {title}
  {Microrheology of semiflexible filament solutions based on relaxation
  simulations},}\ }\href@noop {} {\bibfield  {journal} {\bibinfo  {journal}
  {Soft Matter}\ }\textbf {\bibinfo {volume} {17}},\ \bibinfo {pages} {2920}
  (\bibinfo {year} {2021})}\BibitemShut {NoStop}%
\bibitem [{\citenamefont {Romer}, \citenamefont {Scheffold},\ and\
  \citenamefont {Schurtenberger}(2000)}]{romer2000prl}%
  \BibitemOpen
  \bibfield  {author} {\bibinfo {author} {\bibfnamefont {S.}~\bibnamefont
  {Romer}}, \bibinfo {author} {\bibfnamefont {F.}~\bibnamefont {Scheffold}}, \
  and\ \bibinfo {author} {\bibfnamefont {P.}~\bibnamefont {Schurtenberger}},\
  }\bibfield  {title} {\enquote {\bibinfo {title} {Sol-gel transition of
  concentrated colloidal suspensions},}\ }\href@noop {} {\bibfield  {journal}
  {\bibinfo  {journal} {Phys. Rev. Lett.}\ }\textbf {\bibinfo {volume} {85}},\
  \bibinfo {pages} {4980} (\bibinfo {year} {2000})}\BibitemShut {NoStop}%
\bibitem [{\citenamefont {Veerman}\ \emph {et~al.}(2006)\citenamefont
  {Veerman}, \citenamefont {Rajagopal}, \citenamefont {Palla}, \citenamefont
  {Pochan}, \citenamefont {Schneider},\ and\ \citenamefont
  {Furst}}]{veerman2006macromol}%
  \BibitemOpen
  \bibfield  {author} {\bibinfo {author} {\bibfnamefont {C.}~\bibnamefont
  {Veerman}}, \bibinfo {author} {\bibfnamefont {K.}~\bibnamefont {Rajagopal}},
  \bibinfo {author} {\bibfnamefont {C.~S.}\ \bibnamefont {Palla}}, \bibinfo
  {author} {\bibfnamefont {D.~J.}\ \bibnamefont {Pochan}}, \bibinfo {author}
  {\bibfnamefont {J.~P.}\ \bibnamefont {Schneider}}, \ and\ \bibinfo {author}
  {\bibfnamefont {E.~M.}\ \bibnamefont {Furst}},\ }\bibfield  {title} {\enquote
  {\bibinfo {title} {Gelation kinetics of $\beta$-hairpin peptide hydrogel
  networks},}\ }\href@noop {} {\bibfield  {journal} {\bibinfo  {journal}
  {Macromolecules}\ }\textbf {\bibinfo {volume} {39}},\ \bibinfo {pages}
  {6608--6614} (\bibinfo {year} {2006})}\BibitemShut {NoStop}%
\bibitem [{\citenamefont {Larsen}\ and\ \citenamefont
  {Furst}(2008)}]{furst2008prl}%
  \BibitemOpen
  \bibfield  {author} {\bibinfo {author} {\bibfnamefont {T.~H.}\ \bibnamefont
  {Larsen}}\ and\ \bibinfo {author} {\bibfnamefont {E.~M.}\ \bibnamefont
  {Furst}},\ }\bibfield  {title} {\enquote {\bibinfo {title} {Microrheology of
  the liquid-solid transition during gelation},}\ }\href@noop {} {\bibfield
  {journal} {\bibinfo  {journal} {Phys. Rev. Lett.}\ }\textbf {\bibinfo
  {volume} {100}},\ \bibinfo {pages} {146001} (\bibinfo {year}
  {2008})}\BibitemShut {NoStop}%
\bibitem [{\citenamefont {Larsen}, \citenamefont {Schultz},\ and\ \citenamefont
  {Furst}(2008)}]{larsen2008koreaust}%
  \BibitemOpen
  \bibfield  {author} {\bibinfo {author} {\bibfnamefont {T.~H.}\ \bibnamefont
  {Larsen}}, \bibinfo {author} {\bibfnamefont {K.~M.}\ \bibnamefont {Schultz}},
  \ and\ \bibinfo {author} {\bibfnamefont {E.~M.}\ \bibnamefont {Furst}},\
  }\bibfield  {title} {\enquote {\bibinfo {title} {Hydrogel microrheology near
  the liquid-solid transition},}\ }\href@noop {} {\bibfield  {journal}
  {\bibinfo  {journal} {Korea-Aust. Rheol. J.}\ }\textbf {\bibinfo {volume}
  {20}},\ \bibinfo {pages} {165–173} (\bibinfo {year} {2008})}\BibitemShut
  {NoStop}%
\bibitem [{\citenamefont {Larsen}\ \emph {et~al.}(2009)\citenamefont {Larsen},
  \citenamefont {Branco}, \citenamefont {Rajagopal}, \citenamefont
  {Schneider},\ and\ \citenamefont {Furst}}]{larsen2009macromol}%
  \BibitemOpen
  \bibfield  {author} {\bibinfo {author} {\bibfnamefont {T.~H.}\ \bibnamefont
  {Larsen}}, \bibinfo {author} {\bibfnamefont {M.~C.}\ \bibnamefont {Branco}},
  \bibinfo {author} {\bibfnamefont {K.}~\bibnamefont {Rajagopal}}, \bibinfo
  {author} {\bibfnamefont {J.~P.}\ \bibnamefont {Schneider}}, \ and\ \bibinfo
  {author} {\bibfnamefont {E.~M.}\ \bibnamefont {Furst}},\ }\bibfield  {title}
  {\enquote {\bibinfo {title} {Sequence-dependent gelation kinetics of
  $\beta$-hairpin peptide hydrogels},}\ }\href@noop {} {\bibfield  {journal}
  {\bibinfo  {journal} {Macromolecules}\ }\textbf {\bibinfo {volume} {42}},\
  \bibinfo {pages} {8443–8450} (\bibinfo {year} {2009})}\BibitemShut
  {NoStop}%
\bibitem [{\citenamefont {Corrigan}\ and\ \citenamefont
  {Donald}(2009{\natexlab{a}})}]{donald2009eurphysJE}%
  \BibitemOpen
  \bibfield  {author} {\bibinfo {author} {\bibfnamefont {A.~M.}\ \bibnamefont
  {Corrigan}}\ and\ \bibinfo {author} {\bibfnamefont {A.~M.}\ \bibnamefont
  {Donald}},\ }\bibfield  {title} {\enquote {\bibinfo {title} {Particle
  tracking microrheology of gel-forming amyloid fibril networks},}\ }\href@noop
  {} {\bibfield  {journal} {\bibinfo  {journal} {Eur. Phys. J. E}\ }\textbf
  {\bibinfo {volume} {28}},\ \bibinfo {pages} {457} (\bibinfo {year}
  {2009}{\natexlab{a}})}\BibitemShut {NoStop}%
\bibitem [{\citenamefont {Corrigan}\ and\ \citenamefont
  {Donald}(2009{\natexlab{b}})}]{donald2009langmuir}%
  \BibitemOpen
  \bibfield  {author} {\bibinfo {author} {\bibfnamefont {A.~M.}\ \bibnamefont
  {Corrigan}}\ and\ \bibinfo {author} {\bibfnamefont {A.~M.}\ \bibnamefont
  {Donald}},\ }\bibfield  {title} {\enquote {\bibinfo {title} {Passive
  microrheology of solvent-induced fibrillar protein networks},}\ }\href@noop
  {} {\bibfield  {journal} {\bibinfo  {journal} {Langmuir}\ }\textbf {\bibinfo
  {volume} {25}},\ \bibinfo {pages} {8599--8605} (\bibinfo {year}
  {2009}{\natexlab{b}})}\BibitemShut {NoStop}%
\bibitem [{\citenamefont {Schultz}\ and\ \citenamefont
  {Furst}(2012)}]{schultz2012softmatter}%
  \BibitemOpen
  \bibfield  {author} {\bibinfo {author} {\bibfnamefont {K.~M.}\ \bibnamefont
  {Schultz}}\ and\ \bibinfo {author} {\bibfnamefont {E.~M.}\ \bibnamefont
  {Furst}},\ }\bibfield  {title} {\enquote {\bibinfo {title} {Microrheology of
  biomaterial hydrogelators},}\ }\href@noop {} {\bibfield  {journal} {\bibinfo
  {journal} {Soft Matter}\ }\textbf {\bibinfo {volume} {8}},\ \bibinfo {pages}
  {6198} (\bibinfo {year} {2012})}\BibitemShut {NoStop}%
\bibitem [{\citenamefont {Aufderhorst-Roberts}, \citenamefont {Frith},\ and\
  \citenamefont {Donald}(2012)}]{donald2012softmatter}%
  \BibitemOpen
  \bibfield  {author} {\bibinfo {author} {\bibfnamefont {A.}~\bibnamefont
  {Aufderhorst-Roberts}}, \bibinfo {author} {\bibfnamefont {W.~J.}\
  \bibnamefont {Frith}}, \ and\ \bibinfo {author} {\bibfnamefont {A.~M.}\
  \bibnamefont {Donald}},\ }\bibfield  {title} {\enquote {\bibinfo {title}
  {Micro-scale kinetics and heterogeneity of a p\uppercase{H} triggered
  hydrogel},}\ }\href@noop {} {\bibfield  {journal} {\bibinfo  {journal} {Soft
  Matter}\ }\textbf {\bibinfo {volume} {8}},\ \bibinfo {pages} {5940} (\bibinfo
  {year} {2012})}\BibitemShut {NoStop}%
\bibitem [{\citenamefont {Aufderhorst-Roberts}, \citenamefont {Frith},\ and\
  \citenamefont {Donald}(2014)}]{donald2014eurphysJE}%
  \BibitemOpen
  \bibfield  {author} {\bibinfo {author} {\bibfnamefont {A.}~\bibnamefont
  {Aufderhorst-Roberts}}, \bibinfo {author} {\bibfnamefont {W.~J.}\
  \bibnamefont {Frith}}, \ and\ \bibinfo {author} {\bibfnamefont {A.~M.}\
  \bibnamefont {Donald}},\ }\bibfield  {title} {\enquote {\bibinfo {title} {A
  microrheological study of hydrogel kinetics and micro-heterogeneity},}\
  }\href@noop {} {\bibfield  {journal} {\bibinfo  {journal} {Eur. Phys. J. E}\
  }\textbf {\bibinfo {volume} {37}},\ \bibinfo {pages} {44} (\bibinfo {year}
  {2014})}\BibitemShut {NoStop}%
\bibitem [{\citenamefont {Romer}\ \emph {et~al.}(2014)\citenamefont {Romer},
  \citenamefont {Bissig}, \citenamefont {Schurtenberger},\ and\ \citenamefont
  {Scheffold}}]{romer2014epl}%
  \BibitemOpen
  \bibfield  {author} {\bibinfo {author} {\bibfnamefont {S.}~\bibnamefont
  {Romer}}, \bibinfo {author} {\bibfnamefont {H.}~\bibnamefont {Bissig}},
  \bibinfo {author} {\bibfnamefont {P.}~\bibnamefont {Schurtenberger}}, \ and\
  \bibinfo {author} {\bibfnamefont {F.}~\bibnamefont {Scheffold}},\ }\bibfield
  {title} {\enquote {\bibinfo {title} {Rheology and internal dynamics of
  colloidal gels from the dilute to the concentrated regime},}\ }\href@noop {}
  {\bibfield  {journal} {\bibinfo  {journal} {Europhys. Lett.}\ }\textbf
  {\bibinfo {volume} {108}},\ \bibinfo {pages} {48006} (\bibinfo {year}
  {2014})}\BibitemShut {NoStop}%
\bibitem [{\citenamefont {Havlin}\ and\ \citenamefont
  {Ben-Avraham}(2002)}]{BenAvraham2002}%
  \BibitemOpen
  \bibfield  {author} {\bibinfo {author} {\bibfnamefont {S.}~\bibnamefont
  {Havlin}}\ and\ \bibinfo {author} {\bibfnamefont {D.}~\bibnamefont
  {Ben-Avraham}},\ }\bibfield  {title} {\enquote {\bibinfo {title} {Diffusion
  in disordered media},}\ }\href@noop {} {\bibfield  {journal} {\bibinfo
  {journal} {Adv. Phys.}\ }\textbf {\bibinfo {volume} {51}},\ \bibinfo {pages}
  {187--292} (\bibinfo {year} {2002})}\BibitemShut {NoStop}%
\bibitem [{\citenamefont {Waigh}(2016)}]{Waigh_2016}%
  \BibitemOpen
  \bibfield  {author} {\bibinfo {author} {\bibfnamefont {T.~A.}\ \bibnamefont
  {Waigh}},\ }\bibfield  {title} {\enquote {\bibinfo {title} {Advances in the
  microrheology of complex fluids},}\ }\href@noop {} {\bibfield  {journal}
  {\bibinfo  {journal} {Rep. Prog. Phys. 79}\ }\textbf {\bibinfo {volume}
  {79}},\ \bibinfo {pages} {074601} (\bibinfo {year} {2016})}\BibitemShut
  {NoStop}%
\bibitem [{\citenamefont {Evans}\ \emph {et~al.}(2009)\citenamefont {Evans},
  \citenamefont {Tassieri}, \citenamefont {Auhl},\ and\ \citenamefont
  {Waigh}}]{Evans2009}%
  \BibitemOpen
  \bibfield  {author} {\bibinfo {author} {\bibfnamefont {R.~M.~L.}\
  \bibnamefont {Evans}}, \bibinfo {author} {\bibfnamefont {M.}~\bibnamefont
  {Tassieri}}, \bibinfo {author} {\bibfnamefont {D.}~\bibnamefont {Auhl}}, \
  and\ \bibinfo {author} {\bibfnamefont {T.~A.}\ \bibnamefont {Waigh}},\
  }\bibfield  {title} {\enquote {\bibinfo {title} {Direct conversion of
  rheological compliance measurements into storage and loss moduli},}\
  }\href@noop {} {\bibfield  {journal} {\bibinfo  {journal} {Phys. Rev. E}\
  }\textbf {\bibinfo {volume} {80}},\ \bibinfo {pages} {012501} (\bibinfo
  {year} {2009})}\BibitemShut {NoStop}%
\bibitem [{\citenamefont {Rizzi}(2017)}]{rizzi2017jcp}%
  \BibitemOpen
  \bibfield  {author} {\bibinfo {author} {\bibfnamefont {L.~G.}\ \bibnamefont
  {Rizzi}},\ }\bibfield  {title} {\enquote {\bibinfo {title} {On the
  relationship between the plateau modulus and the threshold frequency in
  peptide gels},}\ }\href@noop {} {\bibfield  {journal} {\bibinfo  {journal}
  {J. Chem. Phys.}\ }\textbf {\bibinfo {volume} {147}},\ \bibinfo {pages}
  {244902} (\bibinfo {year} {2017})}\BibitemShut {NoStop}%
\bibitem [{\citenamefont {Krall}, \citenamefont {Huang},\ and\ \citenamefont
  {Weitz}(1997)}]{KRALL199719}%
  \BibitemOpen
  \bibfield  {author} {\bibinfo {author} {\bibfnamefont {A.}~\bibnamefont
  {Krall}}, \bibinfo {author} {\bibfnamefont {Z.}~\bibnamefont {Huang}}, \ and\
  \bibinfo {author} {\bibfnamefont {D.}~\bibnamefont {Weitz}},\ }\bibfield
  {title} {\enquote {\bibinfo {title} {Dynamics of density fluctuations in
  colloidal gels},}\ }\href@noop {} {\bibfield  {journal} {\bibinfo  {journal}
  {Physica A}\ }\textbf {\bibinfo {volume} {235}},\ \bibinfo {pages} {19--33}
  (\bibinfo {year} {1997})}\BibitemShut {NoStop}%
\bibitem [{\citenamefont {Krall}\ and\ \citenamefont
  {Weitz}(1998)}]{krall1998prl}%
  \BibitemOpen
  \bibfield  {author} {\bibinfo {author} {\bibfnamefont {A.~H.}\ \bibnamefont
  {Krall}}\ and\ \bibinfo {author} {\bibfnamefont {D.~A.}\ \bibnamefont
  {Weitz}},\ }\bibfield  {title} {\enquote {\bibinfo {title} {Internal dynamics
  and elasticity of fractal colloidal gels},}\ }\href@noop {} {\bibfield
  {journal} {\bibinfo  {journal} {Phys. Rev. Lett.}\ }\textbf {\bibinfo
  {volume} {80}},\ \bibinfo {pages} {778} (\bibinfo {year} {1998})}\BibitemShut
  {NoStop}%
\bibitem [{\citenamefont {Teixeira}, \citenamefont {Geissler},\ and\
  \citenamefont {Licinio}(2007)}]{teixeira2007jphyschemB}%
  \BibitemOpen
  \bibfield  {author} {\bibinfo {author} {\bibfnamefont {A.~V.}\ \bibnamefont
  {Teixeira}}, \bibinfo {author} {\bibfnamefont {E.}~\bibnamefont {Geissler}},
  \ and\ \bibinfo {author} {\bibfnamefont {P.}~\bibnamefont {Licinio}},\
  }\bibfield  {title} {\enquote {\bibinfo {title} {Dynamic scaling of polymer
  gels comprising nanoparticles},}\ }\href@noop {} {\bibfield  {journal}
  {\bibinfo  {journal} {J. Phys. Chem. B}\ }\textbf {\bibinfo {volume} {111}},\
  \bibinfo {pages} {340} (\bibinfo {year} {2007})}\BibitemShut {NoStop}%
\bibitem [{\citenamefont {Crooks}(2019)}]{Crooks2019}%
  \BibitemOpen
  \bibfield  {author} {\bibinfo {author} {\bibfnamefont {G.~E.}\ \bibnamefont
  {Crooks}},\ }\href@noop {} {\emph {\bibinfo {title} {Field Guide to
  Continuous Probability Distributions}}}\ (\bibinfo {year} {2019})\BibitemShut
  {NoStop}%
\bibitem [{\citenamefont {Zaccone}\ \emph {et~al.}(2014)\citenamefont
  {Zaccone}, \citenamefont {Winter}, \citenamefont {Siebenb\"urger},\ and\
  \citenamefont {Ballauff}}]{zaccone2014jrheol}%
  \BibitemOpen
  \bibfield  {author} {\bibinfo {author} {\bibfnamefont {A.}~\bibnamefont
  {Zaccone}}, \bibinfo {author} {\bibfnamefont {H.~H.}\ \bibnamefont {Winter}},
  \bibinfo {author} {\bibfnamefont {M.}~\bibnamefont {Siebenb\"urger}}, \ and\
  \bibinfo {author} {\bibfnamefont {M.}~\bibnamefont {Ballauff}},\ }\bibfield
  {title} {\enquote {\bibinfo {title} {Linking self-assembly, rheology, and gel
  transition in attractive colloids},}\ }\href@noop {} {\bibfield  {journal}
  {\bibinfo  {journal} {J. Rheol.}\ }\textbf {\bibinfo {volume} {58}},\
  \bibinfo {pages} {1219--1244} (\bibinfo {year} {2014})}\BibitemShut {NoStop}%
\bibitem [{\citenamefont {Mauro}\ and\ \citenamefont
  {Mauro}(2018)}]{mauro2018physicaA}%
  \BibitemOpen
  \bibfield  {author} {\bibinfo {author} {\bibfnamefont {J.~C.}\ \bibnamefont
  {Mauro}}\ and\ \bibinfo {author} {\bibfnamefont {Y.~Z.}\ \bibnamefont
  {Mauro}},\ }\bibfield  {title} {\enquote {\bibinfo {title} {On the
  \uppercase{P}rony series representation of stretched exponential
  relaxation},}\ }\href@noop {} {\bibfield  {journal} {\bibinfo  {journal}
  {Physica A}\ }\textbf {\bibinfo {volume} {506}},\ \bibinfo {pages} {75--87}
  (\bibinfo {year} {2018})}\BibitemShut {NoStop}%
\bibitem [{\citenamefont {Gradshteyn}\ and\ \citenamefont
  {Ryzhik}(2007)}]{gradshteyn}%
  \BibitemOpen
  \bibfield  {author} {\bibinfo {author} {\bibfnamefont {I.~S.}\ \bibnamefont
  {Gradshteyn}}\ and\ \bibinfo {author} {\bibfnamefont {I.~M.}\ \bibnamefont
  {Ryzhik}},\ }\href@noop {} {\emph {\bibinfo {title} {Table of Integrals,
  Series, and Products}}}\ (\bibinfo  {publisher} {Academic Press, San Diego},\
  \bibinfo {year} {2007})\BibitemShut {NoStop}%
\bibitem [{\citenamefont {C\'ordoba}, \citenamefont {Indei},\ and\
  \citenamefont {Schieber}(2012)}]{cordoba2012jrheol}%
  \BibitemOpen
  \bibfield  {author} {\bibinfo {author} {\bibfnamefont {A.}~\bibnamefont
  {C\'ordoba}}, \bibinfo {author} {\bibfnamefont {T.}~\bibnamefont {Indei}}, \
  and\ \bibinfo {author} {\bibfnamefont {J.~D.}\ \bibnamefont {Schieber}},\
  }\bibfield  {title} {\enquote {\bibinfo {title} {Elimination of inertia from
  a generalized \uppercase{L}angevin equation: Applications to microbead
  rheology modeling and data analysis},}\ }\href@noop {} {\bibfield  {journal}
  {\bibinfo  {journal} {J. Rheol.}\ }\textbf {\bibinfo {volume} {56}},\
  \bibinfo {pages} {185} (\bibinfo {year} {2012})}\BibitemShut {NoStop}%
\bibitem [{\citenamefont {Houghton}, \citenamefont {Hasnain},\ and\
  \citenamefont {Donald}(2008)}]{donald2008eurphysJE}%
  \BibitemOpen
  \bibfield  {author} {\bibinfo {author} {\bibfnamefont {H.~A.}\ \bibnamefont
  {Houghton}}, \bibinfo {author} {\bibfnamefont {I.~A.}\ \bibnamefont
  {Hasnain}}, \ and\ \bibinfo {author} {\bibfnamefont {A.~M.}\ \bibnamefont
  {Donald}},\ }\bibfield  {title} {\enquote {\bibinfo {title} {Particle
  tracking to reveal gelation of hectorite dispersions},}\ }\href@noop {}
  {\bibfield  {journal} {\bibinfo  {journal} {Eur. Phys. J. E}\ }\textbf
  {\bibinfo {volume} {25}},\ \bibinfo {pages} {119} (\bibinfo {year}
  {2008})}\BibitemShut {NoStop}%
\bibitem [{\citenamefont {Rubinstein}\ and\ \citenamefont
  {Colby}(2003)}]{rubinsteinbook}%
  \BibitemOpen
  \bibfield  {author} {\bibinfo {author} {\bibfnamefont {M.}~\bibnamefont
  {Rubinstein}}\ and\ \bibinfo {author} {\bibfnamefont {R.~H.}\ \bibnamefont
  {Colby}},\ }\href@noop {} {\emph {\bibinfo {title} {Polymer Physics}}}\
  (\bibinfo  {publisher} {Oxford University Press},\ \bibinfo {year}
  {2003})\BibitemShut {NoStop}%
\bibitem [{\citenamefont {Bonfanti}\ \emph {et~al.}(2020)\citenamefont
  {Bonfanti}, \citenamefont {Kaplan}, \citenamefont {Charras},\ and\
  \citenamefont {Kabla}}]{bonfanti2020softmatter}%
  \BibitemOpen
  \bibfield  {author} {\bibinfo {author} {\bibfnamefont {A.}~\bibnamefont
  {Bonfanti}}, \bibinfo {author} {\bibfnamefont {J.~L.}\ \bibnamefont
  {Kaplan}}, \bibinfo {author} {\bibfnamefont {G.}~\bibnamefont {Charras}}, \
  and\ \bibinfo {author} {\bibfnamefont {A.}~\bibnamefont {Kabla}},\ }\bibfield
   {title} {\enquote {\bibinfo {title} {Fractional viscoelastic models for
  power-law materials},}\ }\href@noop {} {\bibfield  {journal} {\bibinfo
  {journal} {Soft Matter}\ }\textbf {\bibinfo {volume} {16}},\ \bibinfo {pages}
  {6002} (\bibinfo {year} {2020})}\BibitemShut {NoStop}%
\bibitem [{\citenamefont {Bantawa}\ \emph {et~al.}(2023)\citenamefont
  {Bantawa}, \citenamefont {Keshavarz}, \citenamefont {Geri}, \citenamefont
  {Bouzid}, \citenamefont {Divoux}, \citenamefont {McKinley},\ and\
  \citenamefont {Gado}}]{delgadonatphys}%
  \BibitemOpen
  \bibfield  {author} {\bibinfo {author} {\bibfnamefont {M.}~\bibnamefont
  {Bantawa}}, \bibinfo {author} {\bibfnamefont {B.}~\bibnamefont {Keshavarz}},
  \bibinfo {author} {\bibfnamefont {M.}~\bibnamefont {Geri}}, \bibinfo {author}
  {\bibfnamefont {M.}~\bibnamefont {Bouzid}}, \bibinfo {author} {\bibfnamefont
  {T.}~\bibnamefont {Divoux}}, \bibinfo {author} {\bibfnamefont {G.~H.}\
  \bibnamefont {McKinley}}, \ and\ \bibinfo {author} {\bibfnamefont {E.~D.}\
  \bibnamefont {Gado}},\ }\bibfield  {title} {\enquote {\bibinfo {title} {The
  hidden hierarchical nature of soft particulate gels},}\ }\href@noop {}
  {\bibfield  {journal} {\bibinfo  {journal} {Nature Phys.}\ }\textbf {\bibinfo
  {volume} {19}},\ \bibinfo {pages} {1178--1184} (\bibinfo {year}
  {2023})}\BibitemShut {NoStop}%
\end{thebibliography}

%merlin.mbs aipnum4-1.bst 2010-07-25 4.21a (PWD, AO, DPC) hacked
%Control: key (0)
%Control: author (8) initials jnrlst
%Control: editor formatted (1) identically to author
%Control: production of article title (0) allowed
%Control: page (1) range
%Control: year (1) truncated
%Control: production of eprint (0) enabled
%

\end{document}